\documentclass[oldversion,twocolumns]{aa}
\usepackage{graphicx}

\usepackage{txfonts}
\usepackage{natbib}
\bibpunct{(}{)}{;}{a}{}{,}
\def\na{\ion{Na}{I}}
\def\ca{\ion{Ca}{I}}

\def\kms{$\mbox{km s}^{-1}$}

\def\co{$^{12}$CO\,(2--0)}
\def\dco{$D_{CO}$}
\def\hb{H$\beta$}

\begin{document}

\title{{Integrated K-band spectra of old and intermediate-age globular clusters in the Large Magellanic Cloud}
\thanks{Based on observation collected at the ESO Paranal La Silla
Observatory, Chile, Prog. ID 078.B-0205}}

\titlerunning{Integrated $K$-band spectra of globular clusters in the LMC}

 \author{Mariya Lyubenova
          \inst{1}
          \and
          Harald Kuntschner\inst{2}
	  \and
	  Marina Rejkuba
	  \inst{1}
	  \and
	  David R. Silva
	  \inst{3}
	  \and
	  Markus Kissler-Patig
	  \inst{1}
	  \and
	  Lowell E. Tacconi-Garman
	  \inst{1}
	  \and
	  S{\o}ren S. Larsen
	  \inst{4}
          }

\authorrunning{M.~Lyubenova et al.}

 \institute{ESO, Karl-Schwarzschild-Str. 2, D-85748 Garching bei
 M\"unchen, Germany, \email{mlyubeno@eso.org, mrejkuba@eso.org, mkissler@eso.org, ltacconi@eso.org}
 \and Space Telescope European Coordinating
 Facility, Karl-Schwarzschild-Str. 2, D-85748 Garching bei
 M\"unchen, Germany, \email{hkuntsch@eso.org}
 \and National Optical Astronomy Observatory, 950 North Cherry Ave., Tucson, AZ, 85719
 USA, \email{dsilva@noao.edu}
 \and Astronomical Institute, University of Utrecht, Princetonplein 5, 3584 CC, Utrecht, The Netherlands, 
\email{S.S.Larsen@uu.nl}}

\abstract{Current stellar population models have arguably the largest
  uncertainties in the near-IR wavelength range, partly due to a lack
  of large and well calibrated empirical spectral libraries. In this
  paper we present a project, which aim it is to provide the first
  library of luminosity weighted integrated near-IR spectra of
  globular clusters to be used to test the current stellar population
  models and serve as calibrators for the future ones. Our pilot study
  presents spatially integrated $K$-band spectra of three old
  ($\ge$10~Gyr) and metal poor ([Fe/H]$\sim$~--1.4), and three
  intermediate age (1~--~2~Gyr) and more metal rich
  ([Fe/H]$\sim$~--0.4) globular clusters in the LMC. We measured the
  line strengths of the \na\/, \ca\/ and \co\/ absorption
  features. The \na\/ index decreases with the increasing age and
  decreasing metallicity of the clusters. The \dco\/ index, used to
  measure the \co\/ line strength, is significantly reduced by the
  presence of carbon-rich TP-AGB stars in the globular clusters with
  age $\sim$1~Gyr. This is in contradiction with the predictions of
  the stellar population models of Maraston (2005). We find that this
  disagreement is due to the different CO absorption strength of
  carbon-rich Milky Way TP-AGB stars used in the models and the LMC
  carbon stars in our sample. For globular clusters with age
  $\geq$2~Gyr we find \dco\/ index measurements consistent with the
  model predictions.}


\maketitle

\section{Introduction}
\label{sec:motivation}

Since the 90's, the interpretation of the integrated light of galaxies
(in the nearby universe or at high redshift) heavily relies on
evolutionary population synthesis (EPS) models. Such models were
pioneered by \citet{tinsley80} and the method has been seriously
extended since then
\citep[e.g.][]{bc93,worthey94,vazdekis96,fioc97,starburst99,
  maraston05, schiavon07}. They are used to determine ages, element
abundances, stellar masses, stellar mass functions, etc., of those
stellar populations that are not resolvable into single stars with
today's instrumentation, i.e. most of the universe outside the Local
Group. To build such EPS models we use simple stellar populations
(SSP). There are two essential advantages of focusing on SSPs. First,
SSPs can be reliably calibrated. They can be compared directly with
nearby globular cluster (GC) data for which accurate ages and element
abundances are independently known from studies of the resolved
stars. This step is crucial to fix the stellar population model
parameters that are used to describe model input physics and which
cannot be derived from first principles (e.g., convection, mass loss
and mixing).  Second, SSPs can be used to build more complex stellar
systems. Systems made up by various stellar generations can be
modelled by convolving SSPs with the adopted star formation history
\citep[e.g.][] {kodama97, bc03}. Models describing accurately the
integrated light properties, including medium to high resolution
spectra and/or line-strength indices, are and will be our main tool to
investigate and analyse the star-formation history over cosmological
time-scales.

This approach has worked well in the optical spectroscopic regime and
has led to well calibrated models
\citep[e.g.][]{tmb03,bc03,maraston05}. With the application of such
models to observed spectra we derive reasonable estimates of the main
stellar population parameters (age, chemical composition and M/L
ratio) in the nearby universe
\citep[e.g.,][]{kun00,trager00,thomas05,cappellari06,sb07} as well as
at higher redshifts \citep[e.g.,][]{bernardi05,maraston05,sb09}. Of
course, uncertainties remain due to the degeneracy of age and
metallicity effects in the optical wavelength range \citep[e.g.,][]
{worthey94}.  The integrated near-IR light in stellar populations with
ages $\geq$ 1 Gyr is dominated by one stellar component, cool giant
stars, whose colour and line indices are mainly driven by one
parameter: metallicity \citep{frogel78}.  Near-IR colours and indices
also have the advantage of being more nearly mass-weighted, i.e. the
near-IR mass-to-light ratio is closer to one \citep [see
e.g.,][]{worthey94}. So, by combining the optical as well as near-IR
information one can resolve the currently remaining degeneracies between age and chemical composition, present in the models, and
hope to gain a better understanding of star-formation
histories. However, currently available stellar population models have
arguably the largest uncertainties in the near-IR and thus it is of
paramount importance to provide high-quality observational data to
validate and improve the state-of-the-art models.

Globular clusters in the Local Group are an ideal laboratory for this
project since ample information from studies of the resolved stars is
available. Yet, integrated spectroscopic observations of the Galactic
GCs in the near-IR are very challenging due to their large apparent
sizes on the sky. The Large Magellanic Cloud (LMC) and its globular
cluster system, located about 50 kpc away, is a
much better observational choice. It shows evidence for a very complex
and still ongoing star formation activity. The LMC GCs have an
advantage (for the scope of this project) with respect to Galactic GCs
- they span a larger range in ages. Studies of the LMC globular cluster
system show one old component with age $>$10~Gyr. After this time
there was a ''dark age'' with just one cluster formed before a new
burst of cluster formation that has started around $3-4$~Gyr ago
\citep{dacosta91}. A disadvantage is their lower metallicity.
%
%
\begin{center}
\begin{table*}[tdp]
\centering
\caption{\label{tab:lmc_observations}Target globular clusters in the LMC -- observing log.}
\begin{tabular}{c c c c c c c }
\hline 
\hline 
Name & UT date &  $V$ & $(B-V)$  & SWB & Age & [Fe/H]\\ 
(1) & (2) & (3) & (4) & (5) & (6) & (7) \\
\hline
NGC\,1754 & 2006 Nov 18 & 11.57 & 0.75 & VII & 10 & -1.42$^{a}$, -1.54$^{b}$\\
NGC\,2005 & 2006 Nov 29 & 11.57 & 0.73 & VII & 10 & -1.35$^{a}$, -1.92$^{b}$, -1.80$^{c}$, -1.33$^{d}$\\
NGC\,2019 & 2006 Dec 10 & 10.86 & 0.76 & VII & 10 & -1.23$^{a}$, -1.18$^{b}$, -1.37$^{c}$, -1.10$^{d}$\\
NGC\,1806 & 2006 Dec 06 & 11.10 & 0.73 & V   & 1.1& -0.23$^{b}$, -0.71$^{e}$ \\ 
NGC\,2162 & 2006 Dec 05 & 12.70 & 0.68 & V   &  1.1  & -0.23$^{b}$, -0.46$^{f}$\\ 
NGC\,2173 & 2006 Dec 06 & 11.88 & 0.82 & V-VI & 2& -0.24$^{b}$, -0.42$^{f}$, -0.51$^{g}$\\
\hline
\hline
\end{tabular}

\smallskip 
\flushleft
Notes: (1) Cluster name, (2) date of observation, (3) integrated
$V$-band magnitude, (4) $(B-V)$ colour and (5) SWB type taken from
\citep{bica96,bica99}, (6) Age of the cluster in Gyr, based on the SWB
type \citep{frogel90}, (7) [Fe/H] derived using different methods:
$^{a}$ \citet{olsen98} -- slope of the RGB; $^{b}$ \citet{olszewski91}
-- low-resolution Ca\,II triplet; $^{c}$ \citet{johnson06} --
high-resolution Fe\,I; $^{d}$~\citet{johnson06} -- high-resolution
Fe\,II; $^{e}$ \citet{dirsch00} -- Str\"{o}mgren photometry; $^{f}$
\citet{groch06} -- low-resolution Ca\,II triplet; $^{g}$~\citet{muc08}
-- high-resolution spectroscopy.
\end{table*}
\end{center}
%
%
%
The goal of this project is to provide an empirical near-IR library of
spectra for integrated stellar populations with ages $\geq$ 1 Gyr,
which will be used to verify the predictions of current SSP models in
the near-IR wavelength range. Here we present the results from a pilot
study of $K$-band spectra of 6 globular clusters in the LMC. The
analysis of their $J$ and $H$-band spectra will be discussed in a
separate paper.  This paper is organised as follows: in
Sect.~\ref{sec:sample} we give details about the sample selection
and the observing strategy. Sect.~\ref{sec:obs} is devoted to the
observations and data reduction. In Sect.~\ref{sec:in_or_out} we
discuss the cluster membership of the stars in our sample, in
Sect.~\ref{sec:indices} we describe the near-IR index measurement
procedures. In Sect.~\ref{sec:data_models_lmc} we make a comparison
between the currently available stellar population models in the
near-IR with our data, discuss the observed disagreements, and give
potential explanations. Finally, in Sect.~\ref{sec:conclusions} we
give our concluding remarks.

\section{Sample selection and observational strategy}
\label{sec:sample}

Due to our interests in the application of SSP models to the
integrated light of early-type galaxies, we restricted our sample in
this pilot study to intermediate age (1--2 Gyr) and old ($>$~10 Gyr)
clusters. The sample was selected from the \citet{bica96,bica99}
catalogues by choosing the SWB class \citep{sbw80} to be V, VI or
VII. In this way we ensured that the target systems will have ages
$\ge\,1$\,Gyr. We further required the clusters to be bright
($M_{V}\,<\,-5.8$) and reasonably concentrated (effective radius
$\le\,35\arcsec$). Where no literature data were available, the
concentration was checked by eye on DSS images. Another selection
criterion was the availability of auxiliary data, because we needed
detailed information on age and chemical composition. All of the
selected clusters have HST/WFPC2 and/or ACS imaging
\citep[e.g.][]{olsen98,mackey03}, integrated optical spectroscopy, and
spectra of individual giant stars \citep[e.g.][]{olszewski91,
  beasley02, johnson06, muc08}. We can also benefit from near-IR
studies, both imaging and spectroscopy, of the giant stars in these
clusters \citep [e.g.][]{frogel90,muc06}, as well as of photometry
\citep{persson83,muc06,pessev06}. Taking into account the above
criteria we selected six clusters as targets for this pilot project,
aiming at validating the strategy for observations and analysis (see
Table~\ref{tab:lmc_observations}). Three of the clusters are metal
poor (mean [Fe/H]\,$\sim-1.4$) and have ages of more than 10 Gyr. The
other three are more metal rich (mean [Fe/H]\,$\sim-0.4$) and younger,
with ages between 1 and 2 Gyr. In the literature there are different
age and metallicity estimates for the clusters in our sample,
depending on the methods used. Here we listed the ages based on SWB
types, given in \citet{frogel90} and a compilation of metallicities,
obtained from the literature. A summary of the clusters' properties is
given in Table~\ref{tab:lmc_observations}.

Integrated spectra of clusters with less than $10^{4}\,L_{\sun}$ are
likely to be dominated by statistical fluctuations in the number of
bright AGB and RGB stars \citep[e.g.][for more details see
Sect.~\ref{sec:in_or_out}]{renzini98}. These particular phases of
the stellar evolution are one of the main contributors to the
integrated light of an intermediate age stellar population in the
near-IR \citep[e.g.][]{maraston05}. To sample as much of the cluster
light in a reasonable observing time, we made a mosaic of
3\,$\times$\,3 VLT/SINFONI pointings ($8\arcsec\,\times\,8\arcsec$ per
pointing with a 0\farcs25 spatial sampling) centred on each
cluster. We estimated the total light sampled by the central mosaic
$(24\arcsec\,\times\,24\arcsec)$ for all clusters using the following equation and the 20$\arcsec$ radius aperture $K$-band photometry:

\begin{equation}
\label{eq:puzia}
L_{T}=BC_{K}\cdot 10^{-0.4\cdot (m_{K}-(m-M)-M_{K,\sun} -A_{K})}
\end{equation}
where $m_{K}$ is the observed $K$-band integrated magnitude of the
cluster, $(m-M)=18.5$ is the adopted distance modulus to the LMC
\citep{vdb98,alves02,alves04,borissova04}, $A_{K}$ is the extinction
towards each cluster \citep{zaritsky97}, $M_{K,\sun}=3.33^{m}$ is the
$K$-band absolute magnitude of the Sun \citep{aaq00}. The bolometric
correction $BC_K$ is applied in order to obtain the total bolometric
luminosity $L_{T}$ of the cluster. It depends on the adopted age and
metallicity of the stellar populations. In our case it is 0.6 for the
group of the old and metal poor clusters (NGC\,1754, NGC\,2005 and
NGC\,2019) and 0.3 for the more metal rich and intermediate age
clusters NGC\,1806, NGC\,2162 and NGC\,2173
\citep{maraston98,maraston05}. For all but two clusters we found that
the sampled luminosity is $>\,10^{4}\,L_{\sun}$ (see
Table~\ref{tab:lmc_sampling}). Only the young clusters NGC\,2162 and
NGC\,2173 are at the limit or bellow this value.

In order to maximise the statistical probability of getting the
majority of the RGB and AGB stars, we observed, in addition to the
central mosaics, up to 9 of the brightest stars surrounding each
cluster and outside of the central mosaic. Their selection was based
on $K - (J-K)$ colour-magnitude diagrams from the 2MASS Point Source
Catalogue \citep{2mass} of all the stars with reliable photometry and
located inside the tidal radius of each cluster \citep[$r_t$ taken
from ][see Table~\ref{tab:lmc_sampling}]{mclaughlin05}. We selected
the stars with $(J-K)\,>\,0.9$ and $K>12.5^{m}$ as an initial
separation criterion from the LMC field population.

The inclusion of these additional bright stars in our sample has a
twofold purpose. The original idea, as described above, was to provide
a better sampling of the integrated cluster light by in- or excluding
these bright stars (after a careful decision process, based on
kinematical and chemical composition assumptions). Second, the stars,
which turn out not to be members of any cluster, are representative
for the LMC field star population in the vicinity of our GCs. Thus we
obtained an independent field AGB stars sample for comparison with the
globular clusters. The main properties of these additional stars are
listed in Table~\ref{tab:bs_observations}.

%
%
\begin{center}
\begin{table*}[tdp]
\centering
\caption{\label{tab:lmc_sampling}LMC globular clusters structural and photometric properties.}
\begin{tabular}{c c c c c c c c c c}
\hline 
\hline 
Name      & $r_{h}$    & $r_{t}$ &$K_{r_{t}}$ & $K_{20}$ &
$L_{mosaic}$ & R$_{int}$ & $K_{M06}$ & $(J-K)_{M06}$ & $(H-K)_{M06}$\\ 
(1) & (2) & (3) & (4) & (5) & (6) & (7) & (8) & (9) & (10)\\
\hline
NGC\,1754 & 11\farcs2  &  142\farcs9 & 9.01 & 10.21 & $2.7\times10^{4}$ &  20$\arcsec$ & -- & -- & --\\
NGC\,2005 &  8\farcs65 & 98\farcs8   & 8.92 &  9.80 & $3.9\times10^{4}$ &  20$\arcsec$ & -- & -- & --\\
NGC\,2019 &  9\farcs72 & 121\farcs6  & 8.31 &  9.21 & $6.7\times10^{4}$ &  20$\arcsec$ & -- & -- & --\\
NGC\,1806 &  --        &  --         & 7.57 &  9.19 & $3.4\times10^{4}$ & 100$\arcsec$ & 7.076 & 1.055 & 0.271\\ 
NGC\,2162 & 21\farcs37 & 197\farcs2  & 9.65 & 11.18 & $0.5\times10^{4}$ & 120$\arcsec$ & 9.071 & 1.253 & 0.372\\ 
NGC\,2173 & 34\farcs35 & 393\farcs5  & 8.86 & 10.69 & $0.9\times10^{4}$ &  60$\arcsec$ & 9.050 & 1.033 & 0.297\\
\hline
\hline
\end{tabular}

\smallskip 
\flushleft 
Notes: (1) Cluster name, (2) half-light radius and (3) tidal radius of
the King-model cluster fit from the catalogue of \citet{mclaughlin05},
(4) $K$-band integrated magnitude within the tidal radius from
\citet{pessev06}. For NGC\,1806, where $r_{t}$ is unknown, we list the
magnitude within 200$\arcsec$ radius. The same applies for NGC\,2173,
because aperture equal to the tidal radius does not exist, (5)
$K$-band magnitude from \citet{pessev06} with aperture radius
20$\arcsec$, which matches reasonably well our central SINFONI
mosaics, (6) sampled bolometric luminosity within the clusters central
mosaics in $L_{\sun}$, computed following Equation (\ref{eq:puzia}),
(7) integration radius, up to which we have added stars from the
additional bright AGB stars sample, (8), (9) and (10) photometry with
fixed aperture radius of 90$\arcsec$ of \citet{muc06}.
\end{table*}
\end{center}

\begin{figure*}
\resizebox{\hsize}{!}{\includegraphics[angle=0]{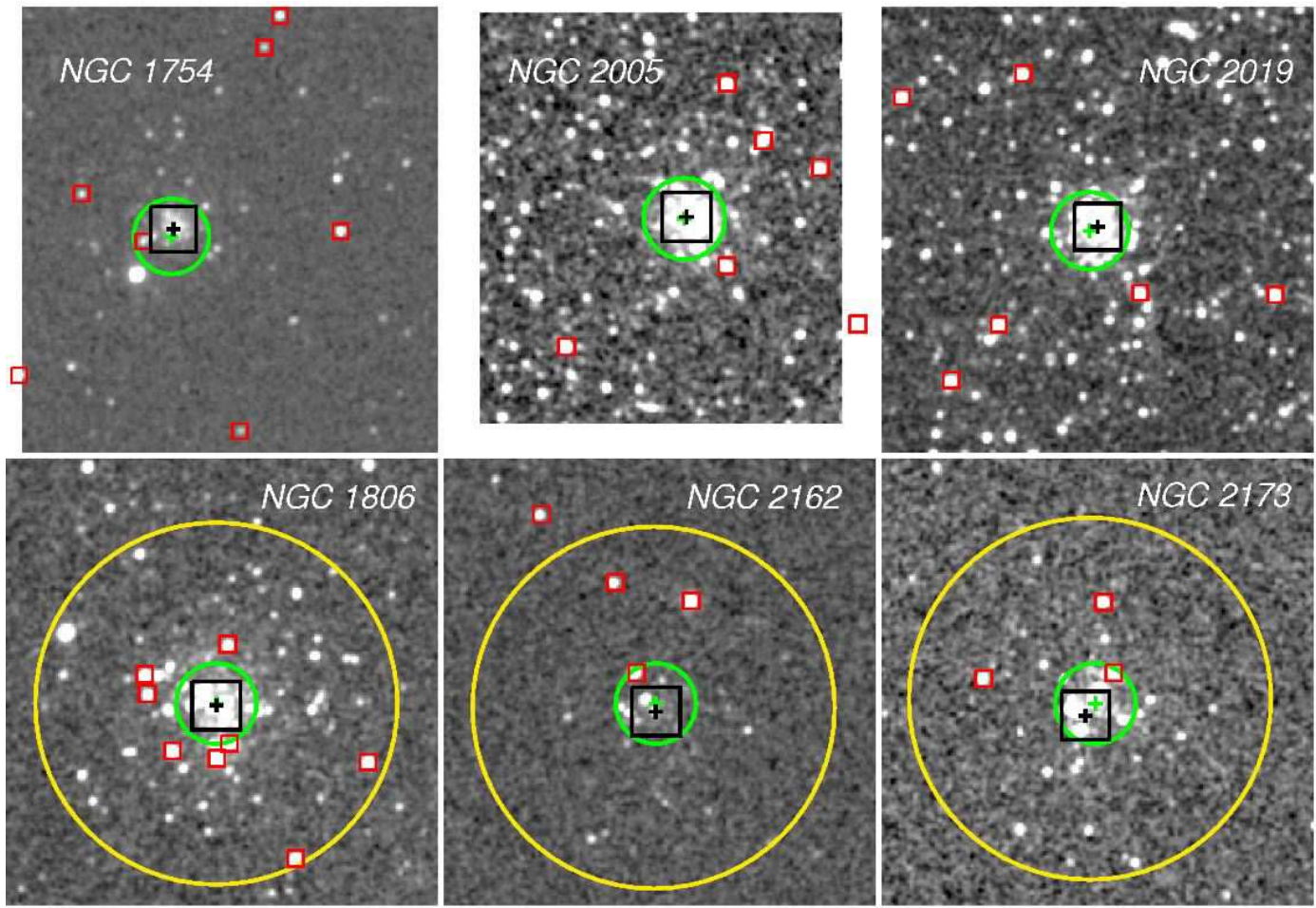}}
\caption{\label{fig:lmc_all_nir} $K$-band 2MASS images of our cluster
  sample.  The black boxes and crosses represent our SINFONI
  $24\arcsec\times24\arcsec$ mosaic FoV and centres, respectively. The
  red squares mark the additional bright stars within the region of
  the 2MASS images that we have observed around each cluster.  The
  green circles and crosses represent the $20\arcsec$ radius aperture
  and the centre used in the photometry by \citet{pessev06}. The large
  yellow circles on the images of the intermediate age clusters (bottom
  row) illustrate the $90\arcsec$ radius aperture of
  \citet{muc06}. The size of each image is
  $3\farcm5\times3\farcm5$. North is up, east -- to the left. The very
  bright star to the south-east of NGC\,1754 was identified as
  galactic foreground star and therefore was not included in the
  analysis. }
\end{figure*}

In Fig.~\ref{fig:lmc_all_nir} we show the 2MASS $K$-band images,
obtained from the 2MASS Extended Source Catalogue of our globular
cluster sample.  The black cross and box on each cluster image match
the centre and the extent of the SINFONI mosaic coverage,
respectively. The red squares mark the additional bright stars,
observed around the clusters within the region of the 2MASS image.
The green circle and cross show the centre and 20$\arcsec$ radius
aperture, which \citet{pessev06} used to obtain integrated magnitudes
for each cluster.  These authors have used near-IR images from the
2MASS Extended Source Catalogue and have performed photometry with
different aperture sizes after correcting for the extinction and LMC
field population.  \citet{muc06} have used ESO 3.5\,m NTT/SOFI images to
provide integrated near-IR magnitudes for half of the globular
clusters in our sample after decontaminating from the LMC field
population and correcting for the extinction and completeness. The
large yellow circles show their 90$\arcsec$ fixed radius aperture.  We
used these photometric studies to compare our spectroscopy with
integrated colours and magnitudes. There is a good agreement between
the centres of our SINFONI observations and the photometry studies,
however, in the case of the sparsely populated cluster NGC\,2173 the
offset is 17\arcsec\/ (in all other cases this offset is smaller than
5\arcsec).

\begin{figure}
\resizebox{\hsize}{!}{\includegraphics[angle=0]{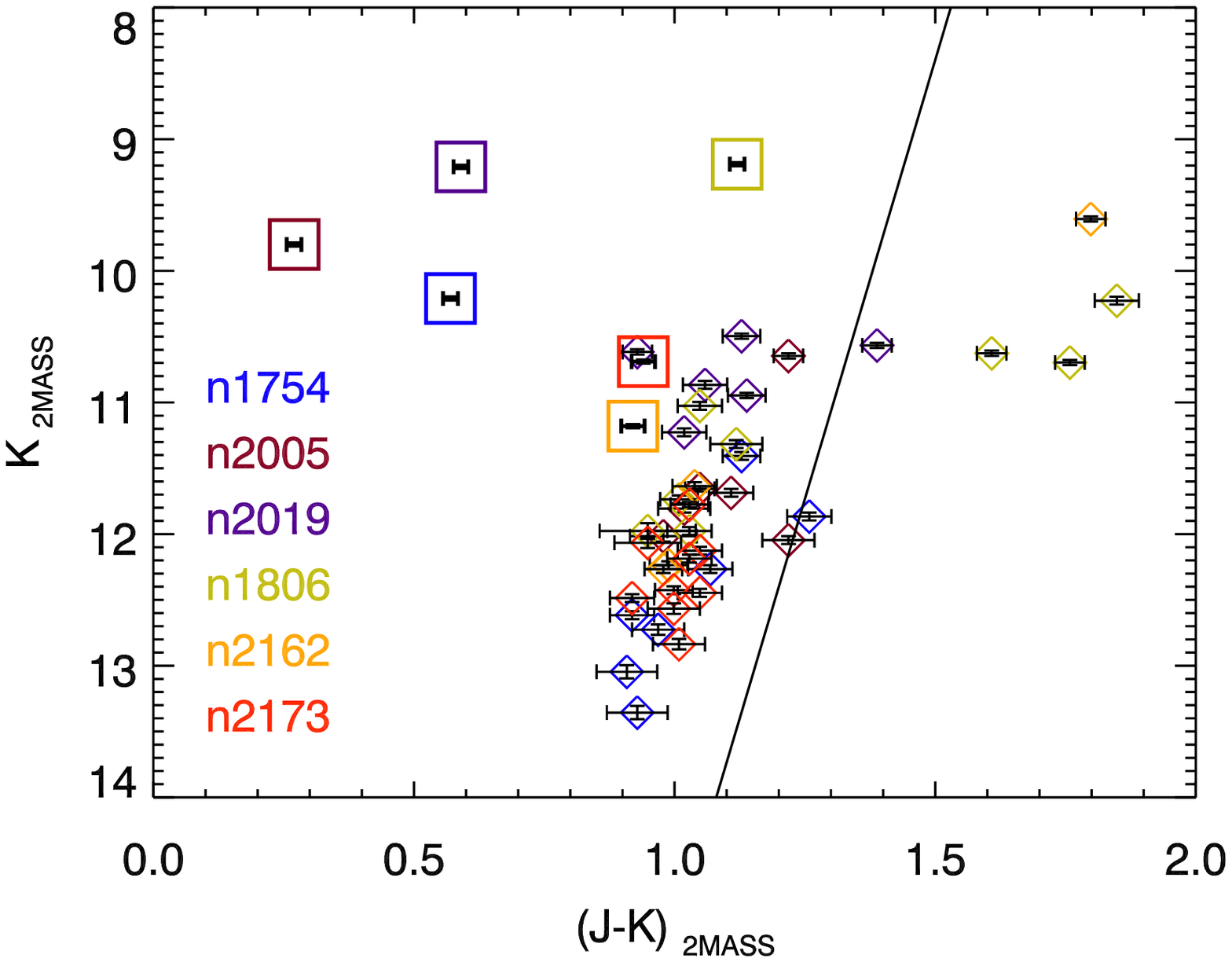}}
\caption{\label{fig:cmd_all} Colour-magnitude diagram including all
  the objects in our observational sample. The photometry of the six
  GCs (coloured square symbols) comes from the catalogue by
  \citet{pessev06}. Data for the additional bright stars (diamond
  symbols) come from the 2MASS Point Source Catalogue
  \citep{2mass}. Colour coding of the symbols for the bright stars
  matches the cluster in whose vicinity they were observed. Stars associated with the old clusters are likely not cluster members as discussed in Sect.~\ref{sec:in_out_old} The slanted line shows the separation between oxygen-rich (leftwards) and carbon-rich (rightwards) stars of \citet{cioni06}.}
\end{figure}

A summary colour-magnitude diagram for all observed objects in our
sample, stars and central mosaics, is shown in
Fig.~\ref{fig:cmd_all}. For the $J$ and $K$-band magnitude of the
central mosaics we adopted the 20$\arcsec$ radius aperture photometry
of \citet{pessev06}, which, as shown on Fig.~\ref{fig:lmc_all_nir},
matches reasonably well our central mosaics.  The selected in addition
bright stars outside the central mosaics are denoted with diamond
symbols. Their photometry comes from the 2MASS Point Source Catalogue
and magnitudes were dereddened following the same method as
\citet{pessev06} -- extinction values were obtained from the
Magellanic Clouds Photometric Survey \citep{zaritsky97} and adopting
the extinction law of \citet{bb88}. The slanted line in this figure
represents the separation between oxygen- and carbon-rich giant stars
of \citet{cioni06}. According to this criterion we have five carbon
rich stars in our sample.

%
%

%
%


\section{Observations and data reduction}
\label{sec:obs}
\subsection{Observations}

The observations of the selected globular clusters and stars were
obtained in service mode in the period October -- December 2006
(Prog. ID 078.B-0205, PI: Kuntschner). We used the integral field unit
spectrograph SINFONI \citep{eis03,bonnet04}, which is mounted in the
Cassegrain focus of Unit Telescope 4 (Yepun) on VLT at Paranal La
Silla Observatory. Its gratings are in the near-IR spectral domain (1
-- 2.5 $\mu$m). We used the $K$-band grating, which covers the
wavelength range from 1.95 to 2.45 $\mu$m at a dispersion of
$2.45\,\AA$/pix. The spectral resolution around the centre of the
wavelength range is R~$\simeq3500$ (6.2~\AA\/ FWHM), as measured from
arc lamp frames.  The largest single pointing with SINFONI covers
$8\arcsec\times8\arcsec$, which is too small even for the most compact
LMC clusters due to their relatively large apparent sizes on the sky.
To sample at least one effective radius of our targets we decided to
use a $3\times3$ mosaic of the largest FoV of SINFONI, thus covering
the central $24\arcsec\times24\arcsec$ for each cluster. Given the
goal to sample the total light, and not to get the best possible
spatial resolution, all the observations were performed in natural
seeing mode, i.e. with no adaptive optics correction. The integration
time was chosen based on the requirement to achieve a signal-to-noise
ratio of at least 50 in the final integrated spectra. The exposure
time for one pointing of the mosaic was 150\,s, divided into three
integrations of 50\,s, dithered by $0\farcs25$ to reject bad
pixels. With the short integration time we could reliably measure and
subtract the very bright and variable near-IR night sky using the data
reduction procedure as described in the next section.

In order to correct for the effects of the night sky, we observed
empty sky regions very close in time and space to our scientific
observations, in a SOOOS sequence for each mosaic pointing (S - sky
integration, O - object).  For the additional bright stars outside the
central mosaic we used the same sequence, but with shorter integration
times of 10\,s per individual integration, leading to a total on
source time of 30\,s.  The sky fields for each cluster were located
outside its tidal radius (see Table~\ref{tab:lmc_sampling}) and were
checked by eye on 2MASS images to be devoid of bright stars. The total
execution time for the longest observing sequence did not exceed 1.5
hours. This ensured that the telluric correction, derived from the
telluric stars, observed after each cluster and additional bright
stars sequence (see Table~\ref{tab:tell_lmc_observations}) would be
sufficiently accurate. The telluric stars were observed at similar
airmass as the clusters.

%
%
\begin{center}
\begin{table*}[tdp]
\centering
\caption{\label{tab:bs_observations}Additional bright RGB and AGB stars.}
\begin{tabular}{c c c c c c c c}
\hline 
\hline 
Name &  R. A. & Dec. & $K$ & $(J-K)$ & \dco & Notes & Cluster \\ 
(1) & (2) & (3) & (4) & (5) & (6) & (7) & (8)\\
\hline
2MASS\,J04540127-7026341 & 04 54 01.27 & -70 26 34.17 & 11.40 & 1.13 & 1.208$\pm$0.013 &   & NGC\,1754 \\
2MASS\,J04540771-7024398 & 04 54 07.72 & -70 24 39.88 & 11.86 & 1.26 & 1.223$\pm$0.021 &   & '' \\
2MASS\,J04543522-7027503 & 04 54 35.22 & -70 27 50.36 & 12.26 & 1.07 & 1.143$\pm$0.015 &   & '' \\
2MASS\,J04542864-7026142 & 04 54 28.64 & -70 26 14.23 & 12.62 & 0.91 & 1.112$\pm$0.018 &   & '' \\
2MASS\,J04540935-7024566 & 04 54 09.35 & -70 24 56.66 & 12.73 & 0.96 & 1.088$\pm$0.021 &   & '' \\
2MASS\,J04541188-7028201 & 04 54 11.88 & -70 28 20.13 & 13.04 & 0.91 & 1.177$\pm$0.030 &   & '' \\
2MASS\,J04540536-7025202 & 04 54 05.36 & -70 25 20.20 & 13.36 & 0.93 & 1.122$\pm$0.029 &   & '' \\
2MASS\,J05302221-6946124 & 05 30 22.21 & -69 46 12.48 & 10.65 & 1.21 & 1.258$\pm$0.008 &   &  NGC\,2005 \\
2MASS\,J05300708-6945327 & 05 30 07.08 & -69 45 32.73 & 11.66 & 1.04 & 1.213$\pm$0.012 &   & '' \\
2MASS\,J05295466-6946014 & 05 29 54.66 & -69 46 01.44 & 11.68 & 1.12 & 1.246$\pm$0.010 &   & '' \\
2MASS\,J05300704-6944031 & 05 30 07.04 & -69 44 03.11 & 11.80 & 1.03 & 1.251$\pm$0.010 &   & '' \\
2MASS\,J05300360-6944311 & 05 30 03.60 & -69 44 31.18 & 12.02 & 0.97 & 1.216$\pm$0.013 &   & '' \\
2MASS\,J05295822-6944445 & 05 29 58.22 & -69 44 44.59 & 12.05 & 1.22 & 1.233$\pm$0.016 &   & '' \\
2MASS\,J05320670-7010248 & 05 32 06.70 & -70 10 24.84 & 10.49 & 1.13 & 1.136$\pm$0.014 &   & NGC\,2019 \\
2MASS\,J05313862-7010093 & 05 31 38.62 & -70 10 09.35 & 10.56 & 1.39 & 1.196$\pm$0.017 & C  & '' \\
2MASS\,J05315232-7010083 & 05 31 52.32 & -70 10 08.39 & 10.61 & 0.94 & 1.200$\pm$0.016 &   & '' \\
2MASS\,J05321152-7010535 & 05 32 11.52 & -70 10 53.52 & 10.87 & 1.06 & 1.176$\pm$0.018 &   & '' \\
2MASS\,J05321647-7008272 & 05 32 16.47 & -70 08 27.26 & 10.95 & 1.13 & 1.183$\pm$0.016 &   & '' \\
2MASS\,J05320418-7008151 & 05 32 04.18 & -70 08 15.15 & 11.22 & 1.02 & 1.204$\pm$0.024 &   & '' \\
2MASS\,J05021232-6759369 & 05 02 12.32 & -67 59 36.92 & 10.23 & 1.84 & 1.051$\pm$0.013 & +,C  & NGC\,1806 \\ 
2MASS\,J05020536-6800266 & 05 02 05.36 & -68 00 26.69 & 10.63 & 1.61 & 1.139$\pm$0.015 &  +,C & '' \\
2MASS\,J05015896-6759387 & 05 01 58.96 & -67 59 38.76 & 10.69 & 1.76 & 1.065$\pm$0.013 &  +,C & '' \\
2MASS\,J05021623-6759332 & 05 02 16.23 & -67 59 33.22 & 11.02 & 1.06 & 1.211$\pm$0.012 &  + & '' \\
2MASS\,J05021870-6758552 & 05 02 18.70 & -67 58 55.20 & 11.32 & 1.11 & 1.267$\pm$0.015 &  + & '' \\
2MASS\,J05021846-6759048 & 05 02 18.46 & -67 59 04.82 & 11.74 & 1.00 & 1.243$\pm$0.018 &   & '' \\
2MASS\,J05021121-6759295 & 05 02 11.21 & -67 59 29.54 & 11.97 & 0.96 & 1.197$\pm$0.026 &   & '' \\
2MASS\,J05021137-6758401 & 05 02 11.37 & -67 58 40.15 & 11.98 & 1.02 & 1.199$\pm$0.023 &   & '' \\
2MASS\,J06002748-6342222 & 06 00 27.48 & -63 42 22.29 &  9.60 & 1.80 & 1.075$\pm$0.012 & +,C  & NGC\,2162 \\
2MASS\,J06003156-6342581 & 06 00 31.56 & -63 42 58.14 & 11.64 & 1.03 & 1.202$\pm$0.011 & +  & '' \\
2MASS\,J06003316-6342131 & 06 00 33.16 & -63 42 13.18 & 12.24 & 0.99 & 1.197$\pm$0.016 & +  & '' \\
2MASS\,J06003869-6341393 & 06 00 38.69 & -63 41 39.37 & 12.26 & 0.99 & 1.184$\pm$0.020 &   & '' \\
2MASS\,J05563892-7258155 & 05 56 38.92 & -72 58 15.52 & 11.77 & 1.03 & 1.227$\pm$0.012 &  & NGC\,2173 \\
2MASS\,J05575667-7258299 & 05 57 56.67 & -72 58 29.91 & 12.07 & 0.95 & 1.166$\pm$0.014 & + & '' \\
2MASS\,J05570233-7257449 & 05 57 02.33 & -72 57 44.95 & 12.13 & 1.04 & 1.159$\pm$0.016 &  & '' \\
2MASS\,J05575784-7257548 & 05 57 57.84 & -72 57 54.89 & 12.18 & 1.03 & 1.203$\pm$0.016 & + & '' \\
2MASS\,J05563368-7257402 & 05 56 33.68 & -72 57 40.28 & 12.43 & 1.00 & 1.146$\pm$0.016 &  & '' \\
2MASS\,J05581142-7258328 & 05 58 11.42 & -72 58 32.87 & 12.45 & 1.04 & 1.154$\pm$0.019 & + & '' \\
2MASS\,J05583257-7258499 & 05 58 32.57 & -72 58 49.90 & 12.48 & 0.92 & 1.136$\pm$0.015 &  & '' \\
2MASS\,J05572334-7256006 & 05 57 23.34 & -72 56 00.66 & 12.56 & 1.01 & 1.227$\pm$0.019 &  & '' \\
2MASS\,J05565761-7254403 & 05 56 57.61 & -72 54 40.38 & 12.84 & 1.01 & 1.169$\pm$0.020 &  & '' \\
\hline
\hline
\end{tabular}

\smallskip 
\flushleft
Notes: (1) 2MASS catalogue star name, (2) and (3) star coordinates (J2000) given in
hours, minutes and seconds, and degrees, arcmin and arcsec, (4) extinction corrected $K$-band magnitude and (5) $(J-K)$ colour from  the 2MASS Point Source Catalogue \citep{2mass}, (6) \dco\/ index value, measured from our spectra, (7) Notes on individual stars: "C" -- a carbon-rich stars, "+" -- the star was used for the integrated spectrum of  the cluster, (8) globular cluster, next to which the star was
observed.
\end{table*}
\end{center}

\subsection{Basic data reduction}

\begin{figure}
\resizebox{\hsize}{!}{\includegraphics[angle=0]{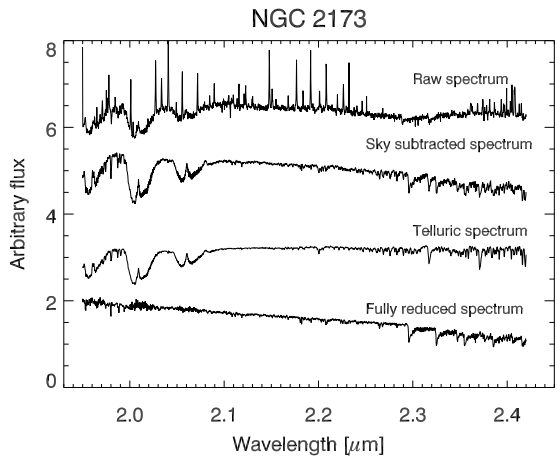}}
\caption{\label{fig:data_red} Overview of the data reduction steps,
applied to obtain fully reduced spectra (in this case, the LMC
globular cluster NGC\,2173). {\it From top to bottom}: (1) Raw
spectrum, (2) Sky subtracted spectrum after running the SINFONI
pipeline. (3) Telluric correction, used to remove the telluric
absorption features. (4) Fully reduced spectrum, used for
analysis. All data reduction steps were performed to the full
data cubes and then, after the telluric correction, integrated spectra
were derived.}
\end{figure}

An overview of the different data reduction steps and their results is
shown in Fig.~\ref{fig:data_red}. There we show a sequence of a raw
spectrum, then the sky subtracted spectrum, the telluric correction
spectrum, and the final, fully reduced spectrum. More details about the data reduction can be found in \citet{phd09}. Here we briefly discuss the most important steps.

The basic data reduction was performed with the ESO SINFONI Pipeline
v. 1.9.2. Calibration products such as distortion maps, flat fields
and bad pixel maps, were obtained with the relevant pipeline tasks
(``recipes''). To reduce the clusters and additional star data, we
divided each observing sequence into cluster frames (27 object plus 10
sky exposures) and star frames (3 star exposures plus one sky per
star). This was needed due to the different integration times for
these two sub-sets.  We also preferred to reduce each mosaic pointing
separately and combine the nine later. In this way we controlled the
quality of each on-source frame sky correction and where needed we
tuned some of the parameters in the pipeline. In summary, we have fed
the {\tt sinfo\_rec\_jitter} recipe with data sets, consisting of one
mosaic pointing (3 on source integrations plus 2 bracketing ``skies'')
and the needed calibration files. Then the recipe extracts the raw
data, applies distortion, bad pixels and flat-field corrections,
wavelength calibration, and stores the combined sky-subtracted spectra
in a 3-dimensional data cube.  The same pipeline steps were also used
to reduce the observations of the additional bright stars and telluric
stars.

The main difficulty during the sky correction arises from the fact
that our observations were sky dominated, combined together with the
small amount of flux in some of the frames. For these cases we found
that by appropriately setting the parameters that indicate the edges
of the object spectrum location ({\tt --skycor-llx, --skycor-lly,
  --skycor-urx} and {\tt --skycor-ury}), we could achieve a good sky
correction.

%
%

\subsection{Telluric corrections}
\label{sec:lmc_tc}

The next data reduction step was to remove the absorption features
originating in the Earth's atmosphere. These features are especially
deep in the blue part of the $K$-band (blue-wards of 2.1\,$\mu$m). For
this purpose we observed after each science target a telluric star,
which is of hotter spectral type (usually A~--~B dwarfs, see
Table~\ref{tab:tell_lmc_observations}). Since these stars are hot
stars, we know that their continuum in the $K$-band is well
approximated by the Rayleigh-Jeans part of the black body spectrum,
associated with their effective temperature. They show only one
prominent feature, the hydrogen Brackett\,$\gamma$ absorption line at
$2.166\,\mu$m. For each telluric star spectrum first we modelled this
line with a Lorentzian profile, with the help of the IRAF task {\tt
  splot}, and then subtracted the model from the star's spectrum. Then
we divided the cleaned star spectrum by a black body spectrum with the
same temperature as the star to remove its continuum shape. Thus we
obtained a normalised telluric spectrum. The last step before applying
it to the science spectra was to scale and shift in the dispersion
direction each telluric spectrum for each data cube by a small amount
($< $ 0.5 pix, 1 pix = 2.45 $\AA$) to minimise the residuals of the
telluric lines (for more details about this procedure see Silva et
al. 2008).  After that each individual cluster mosaic and star data
cube was divided by the so optimised telluric spectrum. In this way we
achieved also a relative flux calibration.

One telluric star, HD\,44533, used for the telluric correction of
NGC\,2019 and its surrounding stars, has an unusual shape of the
Brackett\,$\gamma$ line. It seems to have also some emission together
with the absorption. To remove it, we interpolated linearly the region
between 2.1606 and 2.1706 $\mu$m. In this region there are not many
strong telluric lines, but this interpolation will reflect in an
imperfect correction of the science spectra for this cluster at the
above wavelengths. However, none of the spectral features of interests
for this study reside in this wavelength range.

%
%
\begin{center}
\begin{table}[tdp]
\centering
\caption{\label{tab:tell_lmc_observations}LMC telluric stars observing
log.}
\begin{tabular}{c c c c c}
\hline 
\hline 
Name & Cluster & R. A. & Dec. & SpT \\ 
(1) & (2) & (3) & (4) & (5) \\
\hline
HD\,40624 & NGC\,1754 & 05:54:56.97 & -65:53:00.65 & A0V \\
HD\,43107 & NGC\,2005 & 06:08:44.26 & -68:50:36.27 & B8V \\
HD\,44533 & NGC\,2019 & 06:14:41.94 & -73:37:35.44 & B8V \\
HD\,42525 & NGC\,1806 & 06:06:09.38 & -66:02:22.63 & A0V \\ 
HD\,45796 & NGC\,2162 & 06:24:55.79 & -63:49:41.32 & B6V \\ 
HD\,46668 & NGC\,2173 & 06:27:13.64 & -73:13:42.08 & B8V\\
\hline
\hline
\end{tabular}

\smallskip 
\flushleft
Notes: (1)Telluric star name, (2) Cluster for which this star was
used, (3) and (4) Coordinates of the star, as listed in Simbad (July
2008), (5) Spectral type.
\end{table}
\end{center}

\subsection{Cluster light integration}

\begin{figure}
\resizebox{\hsize}{!}{\includegraphics[angle=0]{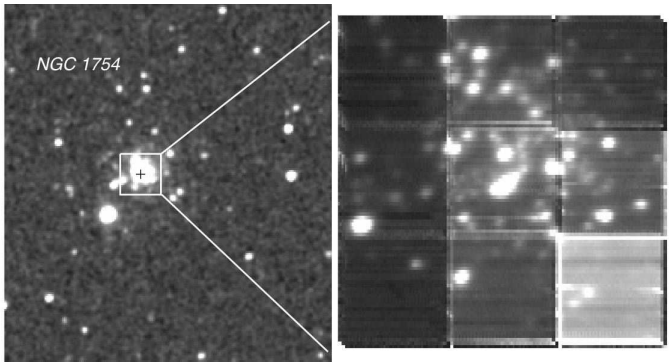}}
\caption{\label{fig:n1754_2mass_sinfo} Two views of NGC\,1754 in the
  $K$-band. The left image is from the 2MASS Extended Source Catalogue
  \citep{2mass}. The white square marks the field, which our SINFONI
  mosaic observations cover ($24\arcsec\times24\arcsec$). The right
  image is reconstructed from the data cube, built up from the
  SINFONI observations. The very bright star to the south-east of the
  central cluster mosaic seen in the left image was identified as a
  galactic foreground star.}
\end{figure}

As a result from the previous data reduction steps we obtained fully
calibrated data cubes for each SINFONI pointing, where the signatures
of the instrument and the night sky are removed as well as
possible. During the next step we reconstructed the full mosaic of
each cluster. For example, Fig.~\ref{fig:n1754_2mass_sinfo} shows
the reconstructed image of the cluster NGC\,1754, together with a
2MASS $K$-band image for comparison. In this image we still see the
imprints of the edges of the individual mosaic tiles. On both images,
the 2MASS and even better on the SINFONI image, we see individual
stars, which can be extracted and studied separately.

However, here we are interested in luminosity weighted, integrated
spectra for each cluster, to compare with stellar population
models. To construct one integrated spectrum per cluster we first
estimated the noise level in each reconstructed image from the mosaic
data cube. We considered this noise is due to residuals after the sky
background correction. Thus we computed the median residual sky noise
level and its standard deviation, after clipping all data points with
intensities more than 3$\sigma$ (assuming that these are the pixels,
which contain the star light from the cluster). After that we selected
all spaxels, which have intensity more than three times the standard
deviation above the median residual sky noise level. We summed them up
and normalised to 1\,s exposure time. In some of the spectra
(NGC\,1754 and NGC\,2019), we still suffered from sky line residuals,
originating from the addition of imperfectly sky corrected spaxels
(this may happen with intrinsically low intensity spaxels). In these
cases we interpolated the contaminated regions.

Our observations were carried out in service mode with constraint sets
allowing seeing up to 2$\arcsec$. For the individual stars this led to
a failure of the standard pipeline recipe while extracting 1D
spectra. Moreover in some cases in the field-of-view there were also
other stars. Thus we decided to manually control the selection of star
light spaxels. We used the same method as for the central clusters
mosaics to obtain the spectra of the additional stars.

\subsection{Error handling}
\label{sec:errors}

The SINFONI pipeline does not provide error estimates, which carry
information about the error propagation during the different data
reduction procedures. Thus we have to use empirical ways of computing
the errors. For this purpose we derived a wavelength dependent
signal-to-noise ratio (S/N) for each cluster integrated spectrum using
the empirical method described by \citet{stoehr07}. We computed the
S/N in 200 pixels width bins from the science spectra (corresponding
to 0.049\,$\mu$m wavelength intervals). Then we fitted a linear
function to the S/N values in the range 2.1\,--\,2.4\,$\mu$m where all
of the spectral features of interest reside. The error spectrum for
each cluster and star was derived by dividing the science spectrum by
the so prepared S/N function. The selection of the bin width was made
after experimenting with a few smaller and larger values. In the case
of bin widths of 50 or 100 pixels (0.01225 or 0.0245\,$\mu$m,
respectively), the S/N function becomes very noisy. Choosing wider,
300 or 400 pixels bins flattens the features of the S/N function. In
general the S/N decreases with increasing wavelength. This is due to
the combination of the spectral energy distribution and the instrument
+ telescope sensitivity. The mean S/N around 2.3\,$\mu$m is larger than 50 for the integrated spectra over the central mosaics of the clusters and in the range 10 -- 40 for the individual stars, depending on their magnitude. The S/N estimate for carbon star spectra is a lower limit due to the numerous absorption features due to carbon based molecules, which are interpreted by the routine as noise. The error spectra were used to estimate the errors of the index measurements, as shown in the plots in this
paper.

%
%

\section{Additional bright AGB stars -- cluster members or not?}
\label{sec:in_or_out}

As discussed in Sect.~\ref{sec:sample} one of the main problems,
when one tries to compile a representative integrated spectrum of a
globular cluster, is the stochastic sampling of bright AGB stars for
clusters with modest luminosities. As an example, \citet{renzini98}
shows that a stellar cluster with $10^{5} L_{\sun}$, solar metallicity
and age of 15\,Gyr will have about $1200$ red giant branch stars
(RGB), $30$ early asymptotic giant branch stars (E-AGB), and only 2
thermally pulsating asymptotic giant branch (TP-AGB) stars.  The
latter stellar evolutionary phase is particularly important  for intermediate age stellar populations of $\sim$~1~Gyr, since up to 80\% of their total $K$-band light
originates there \citep[e.g.][and references therein]{maraston05}. In general the old clusters
(NGC\,1754, NGC\,2005 and NGC\,2019) are massive and well
concentrated, with half-light radii sampled with our SINFONI central
mosaics. Therefore our observations sample $>$~50\% of the total
bolometric light coming from these clusters, and in all cases we
sample at least $\sim3-7\times10^{4}L_{\sun}$.

However, we were not performing that well with the intermediate age
clusters, where for two of them, NGC\,2162 and NGC\,2173, we sample
even less than $10^{4}L_{\sun}$ (see
Table~\ref{tab:lmc_sampling}). Our observations are intrinsically
affected by statistical fluctuations in the number of bright stars,
due to the modest luminosities of the young clusters. These clusters
are less massive and less concentrated than other clusters in our
sample. In the more massive clusters, mass segregation has likely
caused a central concentration of the more massive main sequence
progenitors of the observed AGB stars than the lower mass background
stars that have not yet evolved off the main sequence. In the younger,
less massive clusters, mass segregation has likely been less
efficient. Hence, the AGB progenitors will be less centrally
concentrated relative to other unevolved stars. In turn, this implies
that AGB stars may be found at larger projected radii in lower mass
clusters than in higher mass clusters.

In order to have integrated spectra as representative of our globular clusters, including the most important stellar evolutionary phases, we observed a number of additional bright stars
with near-IR colours and magnitudes in the range expected for RGB and
AGB stars, as explained in the previous sections. After probing their membership, we added some
of them to the central mosaics cluster light to obtain the final integrated spectra, as explained bellow.  We have to note that we have added only integer numbers of bright stars. This is useful when one aims to obtain a representative spectrum for a given globular cluster, what was our goal in this project. In order to achieve an integrated spectrum representative of a full stellar population with a given age and chemical composition, where stochastics does not play a role, one should also consider adding fractional number of bright AGB stars.

The first criterion for cluster membership of the stars is the proximity to the cluster centre, but a LMC field star might also have a relatively nearby position due to projection effects. The
second possibility is to explore the radial velocities of the nearby
stars and the cluster under study. For this purpose we also need to
know the observed velocity dispersions of the stars in the
clusters. Due to the low velocity resolution of our observations we used data from the literature. For the old clusters (NGC\,1754, NGC\,2005 and NGC\,2019) we
took these values from the study of \citet{dubath97}. They have
measured core velocity dispersions from integrated optical spectra,
covering the central $5\arcsec\times5\arcsec$ of each cluster. The
values are 7.\,8$\pm$\,3.0\,\kms\/ for NGC\,1754,
8.1\,$\pm$\,1.3\,\kms\/ for NGC\,2005 and 7.5\,$\pm$\,1.3\,\kms\/ for
NGC\,2019. For the intermediate-age clusters we could not find similar
observed velocity dispersions in the literature, thus we used the
predicted line-of-sight velocity dispersion at the centre of the
cluster by \citet{mclaughlin05}.  The numbers are 1.1\,\kms\/ for
NGC\,2162 and 2.0\,\kms\/ for NGC\,2173. We did not find a similar
estimate for NGC\,1806, so we used a conservative upper limit of
8\,\kms.

\subsection{Old clusters}
\label{sec:in_out_old}

In the case of the three old clusters, NGC\,1754, NGC\,2005 and
NGC\,2019 we cannot exclude any star around any cluster, because their
velocities are consistent with cluster membership. This is due to the
insufficient velocity resolution of our observations. However, some of
the brightest stars that we have observed are located closer to the
tidal radii of the clusters, than to their centres. This is not
expected for old clusters, which have already undergone a core
collapse and have their most massive stars concentrated towards the
centre of the cluster. Indeed, \citet{mackey03} classify the old
clusters in our sample as potential post-core-collapse clusters, due
to the very well expressed power-law cusps in their centres. Moreover,
in Fig.~\ref{fig:cmd_all} the $(J-K)$ colours of the central mosaics
are much bluer than the colours of the additional bright stars around
the old clusters.  \citet{santos97} point out that very young and not massive clusters can be systematically bluer than average due to stochastic sampling of the IMF. However, the integrated spectra of the three old globular clusters discussed here are sampling several times $10^{4}L_{\sun}$ and have ages of $\sim$~10~Gyr. In this regime the simulations of \citet{santos97} predict much less fluctuations in the $(J-K)$ colour than the observed difference between the central mosaics and the additional stars. This additionally led us to the conclusion that these stars are not likely to be cluster members.  In the following we
considered them as members of the LMC field population.  The bright
star marked with a red square just outside the SINFONI field-of-view
for NGC\,1754 in Fig.~\ref{fig:lmc_all_nir} might be a cluster
member, but the S/N of its spectrum is too low ($\sim$\,10) and adding
it to the final integrated cluster spectrum would not increase the
total S/N.

\subsection{Intermediate age clusters}
\label{sec:in_out_young}

This sub-sample includes the clusters NGC\,1806, NGC\,2162 and
NGC\,2173. The last two of them are poorly populated as seen in
near-IR light, visible from Fig.~\ref{fig:lmc_all_nir}, so in their
cases the potential inclusion of additional bright stars to the
central mosaic is very important for the total cluster light
sampling. A detailed photometric study of the RGB and AGB stars in
these clusters is available from \citet{muc06}.  Based on near-IR
colour-magnitude diagrams and after decontamination from the field
population, they report several stars in these evolutionary
phases, as well as their luminosity contribution to the total light of
the clusters. They consider all stars brighter than $K\approx12.3$,
which represents the level of the RGB tip \citep{ferraro04}, and
$(J-K)$ between 0.85 and 2.1 to be AGB stars. 

The AGB stars separate in oxygen rich (M-stars) and carbon rich
(C-stars). An AGB star becomes C-rich, when the amount of dredged up
carbon into the stellar envelope exceeds the amount of oxygen. Then
all the oxygen is bound into CO molecules. The remaining carbon is
used to form CH, CN and C$_{2}$ molecules. This process is more
effective in more metal poor stellar populations, thus they are
expected to have more C-stars \citep{maraston05}. C-stars can
contribute up to 60\% of the total luminosity of metal-poor clusters
\citep{frogel90}. Another important statement that \citet{frogel90}
make is that C and M type stars are found both in clusters and in the
LMC field. Thus it is very difficult to separate intermediate age
globular cluster stars from the field population.

However, the LMC field carbon stars contamination is not expected to
be significant at the locations of the globular clusters in our
sample. The upper limits, according to the carbon stars frequency maps
of \citet{blanco83} are of the order of 0.05 to 0.7 C-type stars for
an area with a radius of 100$\arcsec$.

\smallskip \smallskip {\it NGC\,1806}: This cluster is relatively rich
in stars, as seen on Fig.~\ref{fig:lmc_all_nir}. However, there is a
lack of information about its spectral properties. So far, studies of
the stellar populations have been made mainly with photometric methods
\citep[e.g.][]{frogel90,dirsch00,muc06,mackey08}, and two bright
cluster stars have been observed by \citet{olszewski91} to
spectroscopically estimate the metallicity of the cluster (see
Table~\ref{tab:lmc_observations}). Moreover, there is neither
information about its dynamical properties in the literature, nor
information about its bounds. Thus we had to chose an arbitrary
limiting radius for the bright stars inclusion.

\citet{muc06} point out the presence of 75 RGB stars and 9 AGB stars,
from which 4 are carbon rich stars, in a radius of 90$\arcsec$ from
the centre of the cluster. Indeed, investigating the spectra of the
resolved stars in our SINFONI central mosaic data cube for this
cluster, we identified one of the stars as C-type. The first three of
the brightest additional stars within 90$\arcsec$ are also
C-type. Thus we chose to integrate all the stars within the radius
used by \citet{muc06}, with the exception of the three faintest stars
due to their low S/N spectra.

{\it NGC\,2162}: Inside the aperture photometry radius of 90$\arcsec$,
that \citet{muc06} have used, there is a very bright carbon star with
$K=9.60^{m}$ and $(J-K)=1.80$ (see
Table~\ref{tab:bs_observations}). It is visible in
Fig.~\ref{fig:lmc_all_nir} at $\sim$\,50$\arcsec$ northwest of the cluster centre. If this star is a cluster
member, it would be responsible for $\sim$~60\% of the total $K$-band
cluster light and will affect significantly the integrated spectral
properties of the cluster, thus it is very important to carefully
evaluate its cluster membership. The velocity of this star is within the errors the same as the velocity, which we measured from the central cluster mosaic.
However this is not a definitive proof of its membership, as discussed
above. Looking at the surface distribution maps of C- and M-type
giants across the LMC field in the study of \citet{blanco83}, we see
that NGC\,2162 is located in a region far away from the LMC bar, which
has the highest frequency of field carbon stars. According to these
maps, we can expect to have $\sim$\,25 C-type stars in an area of
1\,deg$^{2}$. This density, scaled to the area, covered by a circle
with a radius of 90$\arcsec$, gives 0.05 carbon stars for the
$1.96\times10^{-3}$\,deg$^{2}$. Moreover, the control field that
\citet{muc06} used to estimate the LMC field contribution (shown on
their Fig. 4), does not contain any stars on the AGB redder than
$(J-K)=1.2$. Having bright carbon stars is not untypical for LMC
globular clusters with the age and metallicity of NGC\,2162, as shown
by \citet{frogel90}. With all these facts taken together, we conclude
that this very bright and red star is most probably a cluster member,
although its definite membership will be confirmed or rejected only by
high resolution spectroscopy. In order to obtain the final integrated
spectrum of NGC\,2162, we have also added two more bright stars,
located within 90$\arcsec$ radius. The remaining stars in our
observational sample are fainter and their spectra have too low
quality. Thus they would not increase the total S/N.

%
%
\begin{figure}
\resizebox{\hsize}{!}{\includegraphics[angle=0]{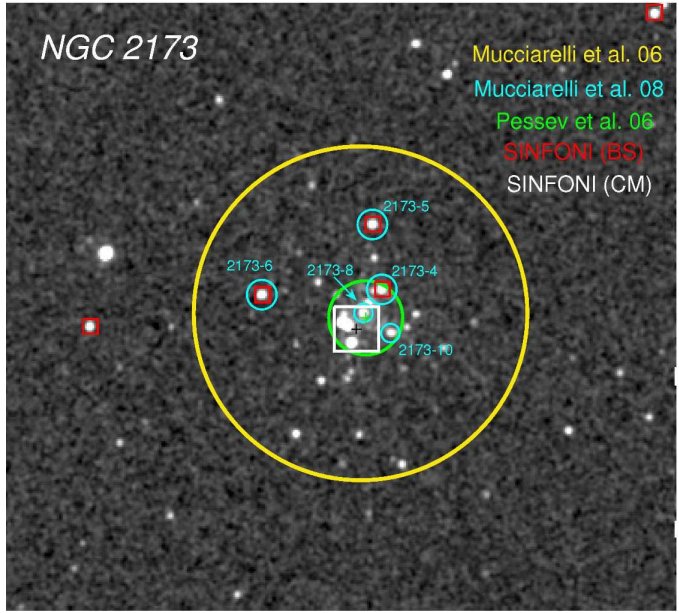}}
\caption{\label{fig:n2173} A summary of the available photometric and
  spectroscopic data about NGC\,2173. The white square marks the
  extent of the SINFONI mosaic. The red squares show the additional
  bright stars we have observed with SINFONI. The cyan circles show
  the stars with high resolution spectroscopy data from \citet{muc08}
  (with numbers assigned as in this paper). The green circle shows the
  aperture with 20$\arcsec$ radius that \citet{pessev06} used for
  photometry. Finally, the yellow circle denotes the photometric
  aperture of \citet{muc06} with 90$\arcsec$ radius.}
\end{figure}

{\it NGC\,2173}: Deciding about the cluster membership of the stars
observed around this cluster was easier, due to the
availability of high resolution spectroscopy of RGB stars in NGC\,2173
from the study of \citet{muc08}. As shown in Fig.~\ref{fig:n2173},
we have three stars outside of and one within the central SINFONI mosaic
 in common with their study. Based on high resolution
spectra \citet{muc08} find that these four stars have very similar
radial velocities (rms = 1.2\,\kms) and negligible star-to-star
scatter in the [Fe/H]. Thus we safely concluded that the first
three stars outside of the SINFONI mosaic are cluster members and we
included them in the final integrated spectrum for the cluster.

Column 7 in Table~\ref{tab:bs_observations} lists the stars, which we added to the final spectra for each globular cluster.

\subsection{Cluster light sampling}
\label{sec:light_sampling}

Our study shows that with a reasonable amount of
mosaic tiles and no more than 1.5 hours VLT observing time, we can sample a significant
fraction of the light from each of our sample of LMC GCs. We have seen that
for well concentrated clusters there is no need for additional light
sampling, other than a central mosaic, covering at least the half-light
radius. The strategy of observing a mosaic on the cluster's centre and
then a sequence of the brightest and closest stars within a certain
radius, works very well in the case of sparse and not luminous
clusters, as well as for those, which are rich, but not well
concentrated. However, there are still some doubts about the cluster
membership of the brightest stars around the intermediate age
clusters, especially in the case of carbon rich stars. For this reason
we would need high resolution spectroscopy to measure their radial
velocities and chemical composition and compare with other RGB stars. 

Flux calibration of ground based spectroscopic data in the near-IR is
particularly difficult, thus we cannot rely on direct luminosity
estimates from our data. However, we can use the available photometry
of \citet{muc06} to estimate the approximate amount of sampled light
in the intermediate age clusters. In this study the authors provide
not only integrated magnitudes and colours, but also estimates of the
M- and C-type stars contributions to the total cluster light in the
$K$-band. From there we know that the $K$-band light of NGC\,2162 is
dominated by only one carbon rich star, which is responsible for
$\sim$\,60\,\% of the total cluster luminosity. This C-type star
contributes with about 70\,\% to our final integrated cluster
spectrum. From this we conclude that for this cluster we are missing
only $\sim$~10\,\% of the luminosity, measured by \citet{muc06}.  The
situation with the other two intermediate age clusters is
similar. NGC\,1806 is quite rich in stars in comparison with the other
two clusters, as seen in the $K$-band images in
Fig.~\ref{fig:lmc_all_nir}. According to \citet{muc06} 22\,\% of the
total cluster light is coming from stars with $K<12.3^{m}$ (AGB
stars). About 77\,\% of it is due to four carbon stars. In our
NGC\,1806 integrated spectrum these four C-type stars account for
$\sim$\,60\,\%. Following \citet{muc06}, 15\,\% of the light in
NGC\,2173 are due to one C-type star. In our spectrum this star is
responsible for $\sim$\,20\,\%. For the three old and metal poor
clusters such detailed estimates of the contributions from different
stellar phases are not available, thus we cannot make similar
estimates for them as for the intermediate age clusters. Our central
SINFONI mosaics cover more than one half-light radius. From this we conclude that the majority of the cluster light is covered.

\begin{figure}
\resizebox{\hsize}{!}{\includegraphics[angle=0]{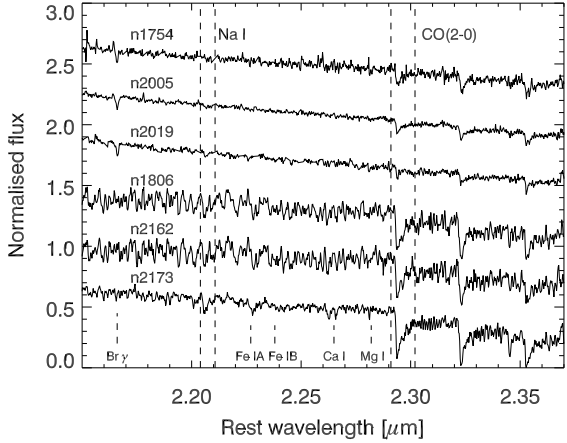}}
\caption{\label{fig:lmc_final_spec} Final integrated spectra of the
  six LMC globular clusters, used for analysis. The spectra of
  NGC\,1806 and NGC\,2162 are clearly dominated by carbon rich stars,
  evident from the numerous carbon based absorption features
  (e.g. C$_{2}$, CH, CN) and the flatter continuum shape.}
\end{figure}

According to these rough estimates it is evident that we have reached
our goal set up in Sect.~\ref{sec:motivation}, namely to obtain
representative luminosity weighted, integrated spectra in the $K$-band
for a group of clusters, having intermediate and old ages.  The final
spectra used for the scientific analysis in the following sections
are shown in Fig.~\ref{fig:lmc_final_spec}.  The achieved
signal-to-noise ratios are given in Table~\ref{tab:lmc_index}. The two
clusters that have spectra dominated by carbon rich giants, NGC\,1806
and NGC\,2162, visually seem to have a significantly lower S/N than the other
clusters. However, as mentioned already in Sect.~\ref{sec:errors}, this does not reflect the
reality: these spectra are dominated by numerous absorption lines from
carbon-based molecules, which the method for computing S/N
interprets as noise. To properly estimate the S/N one would need synthetic spectra,
including the full spectral synthesis. This is clearly not possible,
as we are aiming to provide first templates that could allow
computation of such spectra. We estimated the S/N for these two
clusters based on the amount of integrated flux as compared to the
other clusters in our sample and assigned to them a conservative lower
limit of S/N~$>$~80.

%
%

\section{Line strength indices in the $K$-band}
\label{sec:indices}

The stellar population modelling technique allows us to obtain
detailed estimates of the properties of integrated stellar
populations by measuring, for example, the strengths of selected
absorption or emission features in the spectra and comparing the
measured values to the predicted ones. This approach has shown to be
very effective in the optical wavelength range and a well established
system, the so called Lick/IDS system \citep{faber85,wort94,trager98},
is widely used. For the study of $K$-band spectral features several
index definitions have been adopted \citep[e.g.][]{natasha00,frog01,silva2008,esther08}. To measure the
line strengths of \na\/ and \ca\/ (see Fig.~\ref{fig:lmc_final_spec}) we used the index definitions of
\citet{frog01}, and for the strength of \co\/ the definitions of
\citet{esther08} and \citet{maraston05}.

The principle of measuring a near-IR index is the same as in the
optical Lick system. The value is computed as the ratio of the flux in
a central passband to the flux at the continuum level, measured in two
pseudo-continuum passbands on both sides of the central passband. Due
to the lack of a well defined continuum on the red side of the \co\/
feature, the \dco\/ index that \citet{esther08} defined, uses two
continuum passbands on the blue side. This index measure the ratio
between the average fluxes in the continuum and the absorption
bands. \citet{maraston05} uses a CO index, which measures the ratio
of the flux densities at 2.37 and 2.22\,$\mu$m, based on the
HST/NICMOS filters F237M and F222M. The index is computed in units of
magnitudes and is normalised to Vega. This index reflects the strength
not only of \co\/, but also of the other CO absorption features in the
range 2.3~--~2.4~$\mu$m (see Fig.~\ref{fig:lmc_final_spec}).

Prior to index measurements we broadened our spectra to a spectral
resolution of 6.9~\AA\/ (FWHM) to match the resolution of similar
earlier studies of the $K$-band light of early type galaxies and stars
obtained with VLT/ISAAC \citep{silva2008,esther09}. We measured the
recession velocities of the LMC globular clusters and stars with the
IRAF task {\tt fxcor} and corrected for them. We did not apply
velocity dispersion corrections to the indices due to the very low
velocity dispersions of the clusters ($<$~10~\kms\/, see previous
section). The final index values are listed in
Table~\ref{tab:lmc_index}.

The current stellar population models, which include the near-IR
wavelength range \citep[e.g.][]{maraston05}, have too low spectral
resolution at these wavelengths ($\sim$~200~\AA\/ FWHM) to be able to
give predictions for the weak and narrow features \na\/ and
\ca\/. Until more detailed models become available we can only
empirically explore their dependance on the parameters of the globular
clusters, derived from resolved light studies.

\begin{figure}
\centering
\resizebox{\hsize}{!}{\includegraphics[angle=0]{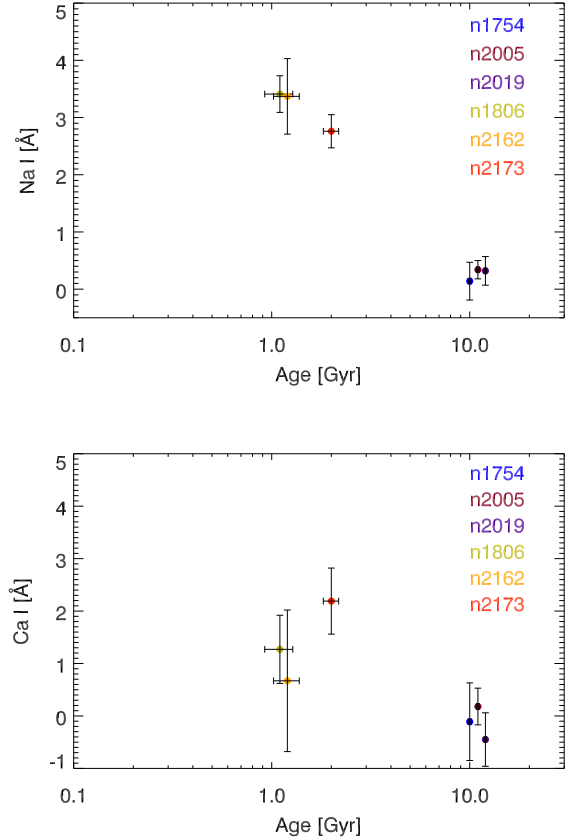}}
\caption{\label{fig:na_ca_age} Near-IR index measurements plotted
  vs. the age of the clusters. {\it Top panel}: \na\/, {\it Bottom
    panel}: \ca\/ index. }
\end{figure}

In Fig.~\ref{fig:na_ca_age} we show the dependance of the \na\/ and
\ca\/ index on the age of the clusters. We see that the
\na\/ index increases with decreasing age of the cluster and
increasing metallicity. This result was already suggested by
\citet{silva2008}, who find that the centres of early type galaxies in
the Fornax cluster with signatures of a recent star formation,
i.e. stronger \hb\/, have also stronger \na\/ indices. The \ca\/ index
seems to show similar behaviour as  \dco\/ (discussed further in the text) as function of age, albeit
with larger error bars.

The \co\/ absorption feature at 2.29~$\mu$m, which we described with
the \dco\/ index, is much stronger and broader. Stellar population
model predictions for its strength exist and will be discussed in the
following section in more detail.

%
\begin{center}
\begin{table*}[tdp]
\centering
\caption{\label{tab:lmc_index}$K$-band indices in LMC globular clusters}
\begin{tabular}{c c c c c c}
\hline 
\hline 
Name & \na\/~[\AA]&  \ca\/~[\AA] & \dco  & CO [mag] & S/N\\ 
(1) & (2) & (3) & (4) & (5) & (6) \\
\hline
NGC\,1754 & 0.14$\pm$0.33 & -0.11$\pm$0.74 & 1.082$\pm$0.005 & -0.035$\pm$0.002 & 55\\
NGC\,2005 & 0.34$\pm$0.16 & 0.18$\pm$0.35 & 1.086$\pm$0.003 & 0.017$\pm$0.001 & 107\\
NGC\,2019 & 0.32$\pm$0.25 & -0.45$\pm$0.51 & 1.068$\pm$0.003 & -0.003$\pm$0.001 & 103\\
NGC\,1806 & 3.41$\pm$0.32 & 1.27$\pm$0.65 & 1.129$\pm$0.005 & 0.026$\pm$0.001 & $>$80\\ 
NGC\,2162 & 3.37$\pm$0.66 & 0.67$\pm$1.35 & 1.108$\pm$0.010 & 0.013$\pm$0.002 & $>$80\\
NGC\,2173 & 2.76$\pm$0.29 & 2.19$\pm$0.63 & 1.186$\pm$0.005 & 0.123$\pm$0.002 & 75\\
\hline
\hline
\end{tabular}

\smallskip 
Notes: (1) Cluster name, (2) \na\/ index, (3) \ca\/ index, (4) \dco\/ index, (5) CO index, (6) signal-to-noise ratio of the integrated spectrum. 
\end{table*}
\end{center}

%
%

\section{Comparisons with stellar population models}
\label{sec:data_models_lmc}

One of the goals of this project was to verify the predictions of the
current stellar population models in the near-IR wavelength range.
Such models are the ones presented in \citet{maraston05}. The author
provides a full set of SEDs (spectral energy distributions) for
different stellar populations with ages from 10$^{3}$\,yr to 15\,Gyr
and covering a range of metallicities. These SEDs extend up to
2.5\,$\mu$m, but their spectral resolution in the near-IR wavelengths
is rather low, with one pixel covering 100\,$\AA$.

The model predictions include the CO index, and for completeness we
measured the \dco\/ index as defined by \citet{esther08}.  In order
to measure it for the models we had to interpolate the model SEDs
linearly to a smaller wavelength step of 14\,$\AA$. We have chosen
this value after a few tests to check the stability of the
computation. \citet{esther08} show that their \dco\/ index definition
is very little dependent on instrumental or internal velocity
broadening of the spectrum. However, the most extreme resolution they
have tested is $\sim$\,70\,$\AA$ ($FWHM$), while the resolution of the
model SEDs is $\sim$~200\,$\AA$ ($FWHM$). We have measured the \dco\/
index in 10 stars from our sample with their nominal resolution and
when broadened to match the resolution of the models. The broadened
spectra have a \dco\/ index value, which is weaker by 0.04 with
respect to the unbroadened spectra. An offset with this size is not
going to influence our general conclusions about the integrated
spectra of LMC GCs. This is further supported by the CO index values
of the GCs, which exhibit the same relative values compared to the
models. The CO index covers a very large wavelength range and is
little dependent on the spectral resolution. Thus we decided to
measure the \dco\/ index at the nominal resolution of our spectra
(6.9~\AA\/ FWHM) and treat the model predictions with caution. Once
higher resolution models become available, this comparison should be
repeated in a more quantitative way. In the following subsection we
discuss the comparison between the models and the data. For clarity we
divided the globular clusters in two groups. In
Sect.~\ref{sec:comparisons_old} we explore the three old and metal
poor clusters NGC\,1754, NGC\,2005 and NGC\,2019. In
Sect.~\ref{sec:carbon_intermediate} we discuss the intermediate age
globular clusters NGC\,1806, NGC\,2162 and NGC\,2173.

\subsection{Old clusters}
\label{sec:comparisons_old}

Literature data of the integrated near-IR colours of our sample of old
and metal poor clusters are shown in
Fig.~\ref{fig:lmc_model_colours} with open circles. The large open
symbols represent the colours, derived from 100$\arcsec$ radius
aperture photometry, the small open symbols from 20$\arcsec$ radius
aperture photometry from \citet{pessev06}. With blue lines we have
overplotted SSP model predictions from \citet{maraston05} for a Kroupa
IMF and blue horizontal branch morphology. The metallicity which is
closest to the one, derived for our GCs sample is denoted with a solid
blue line, [Z/H]\,=\,--1.35. In the top panel the data agree
reasonably well with the model $(J-K)$ colour. In the bottom panel the
$(H-K)$ colours show a larger spread.

In Fig.~\ref{fig:co_models_comp} we show the comparison between
model predictions and index measurements for \co. For the top panel we used the \citet{maraston05} CO index, while for the bottom panel we used the \dco\/ index defined in
\citet{esther08}. For stellar populations with an age of more than
3\,Gyr the near-IR $K$-band light is dominated by RGB stars, whose
contribution stays approximately constant over large time scales. This
is reflected in the stellar population models in
Fig.~\ref{fig:co_models_comp}, where the CO index remains almost
constant at a given metallicity for ages $\ge$\,3 Gyr.  Our
observational data fit reasonably well with the model predictions,
despite the slightly lower index values in the LMC GCs with respect to
the [Z/H] = --\,1.35 line. A possible reason for this, including the
bluer colours of the globular clusters with respect to the models, may
arise from the fact that the models use [Z/H] to describe the
metallicity, which includes not only iron but also other heavy
elements. The literature data, which we used, estimate the
metallicity only based on [Fe/H]. If the globular clusters in our
sample follow similar chemical trends as the old LMC globular clusters
discussed in \citet{muc09} we might expect a better agreement.

\subsection{Intermediate age clusters}
\label{sec:carbon_intermediate}

We divided the intermediate age GCs sample into two sub-samples,
according to the SWB type of the clusters. NGC\,1806 and NGC\,2162
have SWB type V and thus their age is estimated to be around 1.1\,Gyr
\citep{frogel90}. NGC\,2173 has SWB between V and VI and an age of
approximately 2\,Gyr \citep{frogel90}. Based on colour-magnitude
diagram methods different authors give slightly higher or lower ages
for these clusters. In order to be consistent with the age calibration
of the SSP models presented by \citet{maraston05} we have decided to
use the ages based on SWB types, as was done in these models.

\begin{figure}
\resizebox{\hsize}{!}{\includegraphics[angle=0]{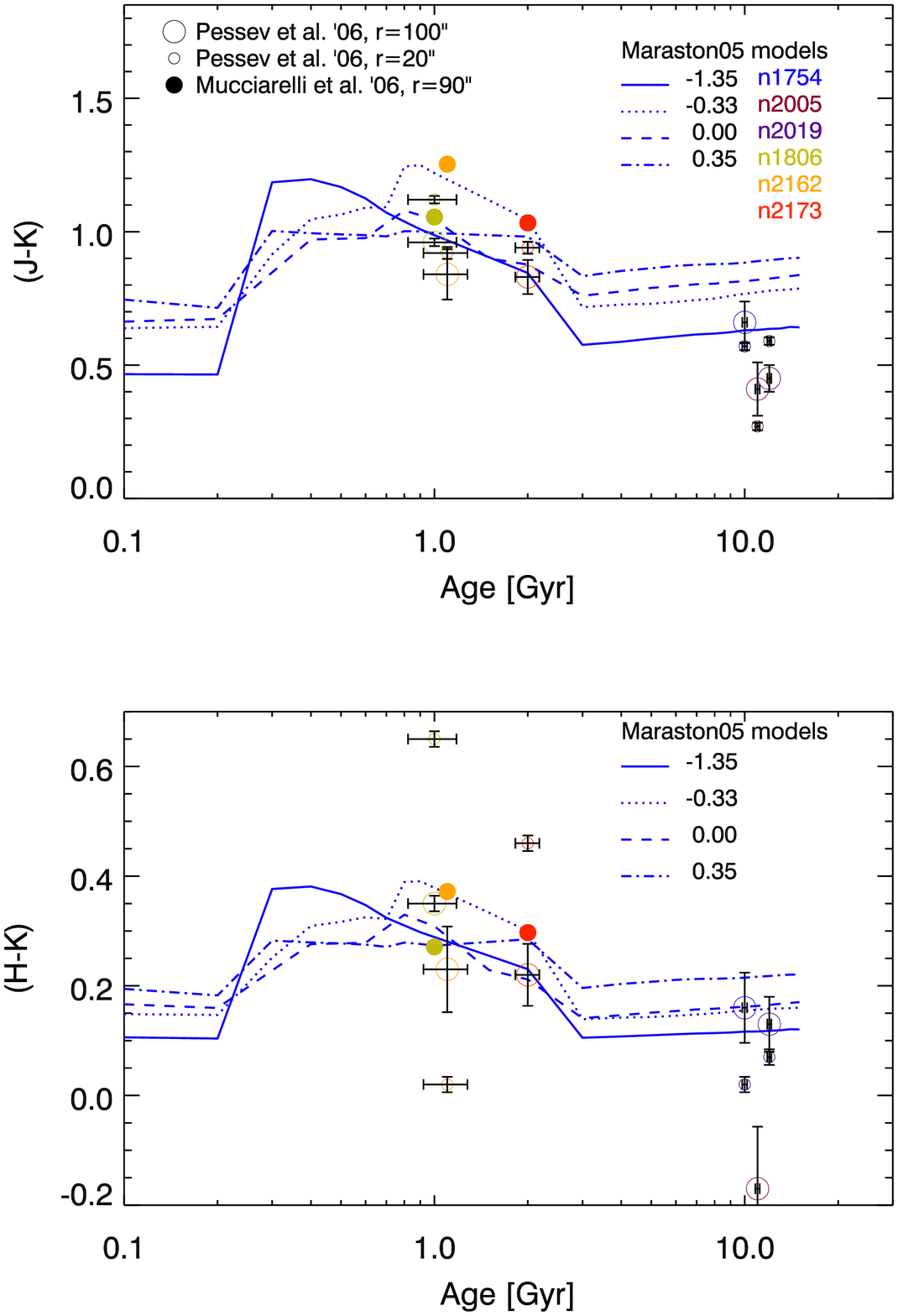}}
\caption{\label{fig:lmc_model_colours} Comparison of clusters'
  integrated colours with SSP model predictions from
  \citet{maraston05}. The filled coloured circles (colour coding as
  listed in the upper right corner of the top plot) correspond to the
  $(J-K)$ and $(H-K)$ colours derived form the 90$\arcsec$ radius
  aperture photometry of \citet{muc06}. The coloured open circles give
  the values, which \citet{pessev06} derived from two different size
  apertures: small circles for $r=20\arcsec$, large circles for
  $r=100\arcsec$. The blue thick lines with different styles give the
  model predictions as a function of the metallicity and the age.}
\end{figure}

In Fig.~\ref{fig:lmc_model_colours}, a compilation of literature
photometric data for the intermediate age GCs in our sample are
shown. According to their mean metallicity the closest model is shown
with the dotted lines and has [Z/H] = --\,0.33. The integrated near-IR
colours within an aperture with radius of 90$\arcsec$ from
\citet{muc06} are shown with solid circles.  The photometry of
\citet{pessev06} with different aperture sizes is shown with open
circles.

Comparing the large filled and open symbols in
Fig.~\ref{fig:lmc_model_colours}, we see evidence for disagreement
between the two studies, despite the fact that \citet{muc06} do not provide
error bars. We also see that according to the different apertures of
\citet{pessev06} the intermediate age clusters exhibit colour
gradients of about 0.2$^{m}$ in $(J-K)$ and $(H-K)$. However, not
always in the same direction. In $(J-K)$ the clusters become
systematically bluer with increasing radius, while in $(H-K)$
NGC\,2162 gets redder. In general colour gradients in globular
clusters can be explained by mass segregation, which makes the most
massive, and thus the most evolved, stars to be concentrated towards
the centre. However, this effect is not expected to be that large, in
comparison with the values that \citet{pessev06} report. The $K$-band magnitudes of \citet{pessev06} are fainter by approximately one magnitude as compared to the ones measured by \citet{muc06} for
NGC\,1806 and NGC\,2162 for a similar aperture and the $(J-K)$ colours
are bluer. These effects could be either due to an overestimated LMC
field decontamination, if \citet{pessev06} removed the reddest stars,
or underestimation of the field in the case of \citet{muc06}. We can
check this hypothesis using NGC\,2162, where the $K$-band light of the
globular cluster is dominated by one single carbon rich star. By
adding its $K$-band magnitude, taken from the 2MASS catalogue, to the
integrated (over 100\arcsec) magnitude of the cluster, taken from
\citet{pessev06}, we obtain a final $K$-band magnitude in much better
agreement with \citet{muc06}. Field carbon stars are not observed
frequently at the location of this cluster in the LMC, thus there is a
high probability that this star is a cluster member. Therefore the
large colour gradients present in the work of \citet{pessev06} and
their fainter $K$-band magnitudes are most probably due to their
oversubtraction of the LMC field star contribution for these two
clusters.

%
%

%
%
\begin{figure}
\resizebox{\hsize}{!}{\includegraphics[angle=0]{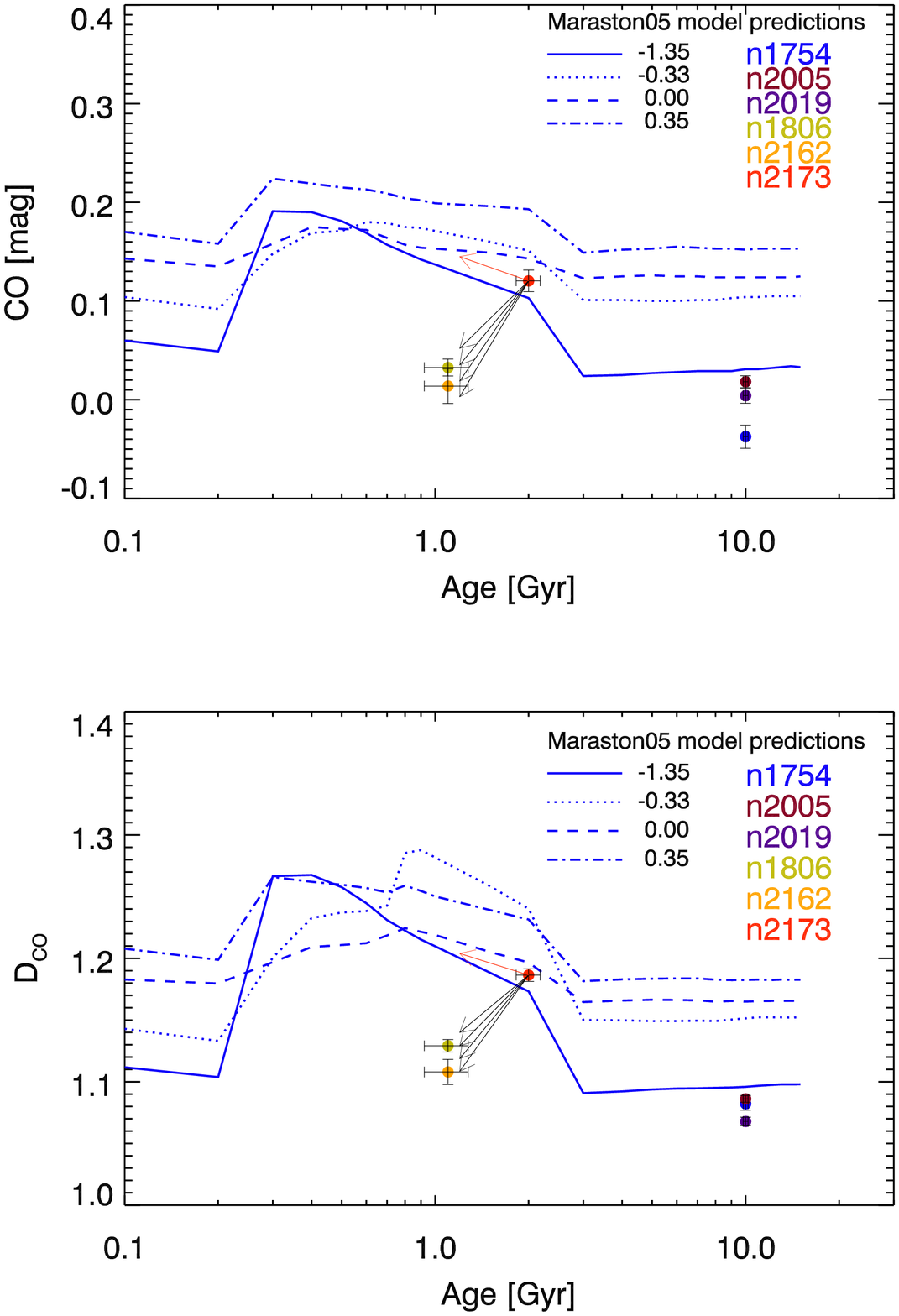}}
\caption{\label{fig:co_models_comp} A comparison of model predictions
  of \citet{maraston05} with our LMC GCs data (coloured filled
  circles). Model metallicities are given with different line styles
  {\it Top panel:} CO index, used by \citet{maraston05} to compute CO
  line strength. {\it Bottom panel:} $D_{CO}$, defined by
  \citet{esther08}. On both plots the arrows show the changes in CO
  after adding a different amount of carbon star contribution to the
  NGC\,2173 spectrum. From top to bottom the carbon star fraction is
  40\,\%, 50\,\%, 60\,\% and 70\,\% respectively. The red arrow shows
  the CO index when the spectrum of NGC\,2173 is composed of 60\,\% of
  the spectrum in the 3$^{rd}$ bin of \citet{lancon02}.}
\end{figure}

Our comparison of the SSP model predictions with the observed CO
indices of LMC globular clusters is shown in
Fig.~\ref{fig:co_models_comp}.  For the intermediate age clusters on
both panels the closest model metallicity to the data is denoted with
the dotted line, [Z/H]= --\,0.33. NGC\,2173 (age $\sim$2~Gyr) is
marked with a red filled circle and agrees reasonably well with the
model predictions. However, the other two intermediate age clusters,
NGC\,1806 (light green filled circle) and NGC\,2162 (orange filled
circle) have CO index values much lower than the model predictions,
independent from the CO index definition.

\subsubsection{Cluster light sampling and stochastic effects}

We have tested a few scenarios, which can provide possible
explanations of the observed trends. The first scenario, which we have
investigated is whether we are sampling well the total cluster
light. We consider this is not a problem following the estimations
about the sampled cluster light during our observations, made in
Sect.~\ref{sec:light_sampling}. The second question is, are the two
clusters stochastically sampling the IMF well? Here the answer is
"most likely".  \citet{lancon00} show that the CO index is significantly influenced by stochastic fluctuations for intermediate age stellar populations with $10^{3} M_{\sun}$, but when the mass reaches $10^{4} M_{\sun}$ and above the fluctuations are much less prominent. Our intermediate age globular clusters have masses closer to the $\sim10^{4} M_{\sun}$ range \citep[e.g. NGC\,2162,][]{mclaughlin05} and thus are not expected to exhibit large variations in the CO index. By building a super cluster from NGC\,1806 and NGC\,2162 we are reducing the stochastic effects further, and we still find the same behaviour of decreasing \dco\/
index value. However, this finding needs to be further verified with
larger samples of intermediate age clusters.

\subsubsection{Different Carbon stars in LMC and Milky Way?}

\begin{figure}
\resizebox{\hsize}{!}{\includegraphics[angle=0]{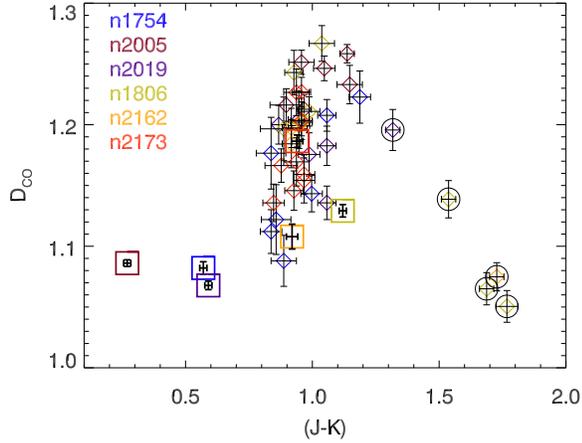}}
\caption{\label{fig:jk_co} $(J-K)$ vs. \dco\/ index of the globular
  clusters and stars in our sample. The squares stand for the central
  mosaics of the GCs, the diamonds show the additional bright stars in
  our sample. The stars denoted with circles are carbon-rich stars.}
\end{figure}

The careful investigation of the integrated spectra of LMC
intermediate age GCs showed us that the presence of C-type stars does
not only influence the overall colours of the clusters, as discussed
above, but also their CO index. Fig.~\ref{fig:jk_co}, where we
plotted the \dco\/ index as a function of the $(J-K)$ colour for all
of the objects in our observational sample, shows a general trend of
increasing \co\/ index with the colour becoming redder. This is
typical for oxygen rich (or M-type) stars \citep[e.g.][]
{frogel90}. However, stars with $(J-K)>$\,1.3 show a decreasing \co\/
line strength. These are exactly the carbon rich stars, separated in
Fig.~\ref{fig:cmd_all} with a slanted line, and denoted in
Fig.~\ref{fig:jk_co} with circles. Similar trends for carbon stars of LMC globular
clusters have been observed by \citet{frogel90}.

%

According to \citet{frogel90} the clusters, which harbour the
brightest carbon stars are of SWB type V and VI. For these clusters,
carbon stars contribute about 40\,\% of the total bolometric light and
this observation is taken into count in the calibrations of the
computed SEDs of \citet{maraston05} models. Using the spectrum of
NGC\,2173, which is very convenient for this kind of tests, because it
is the one with a minimal carbon star contribution in our sample, we
checked what is the effect of different ratios of M-type to C-type
stars on the CO line strength of the final integrated spectrum. The
results are shown in Fig.~\ref{fig:co_models_comp} with black
arrows, which start from the initial position of NGC 2173 (red filled
circle) and end at the age of clusters with SWB type V,
i.e. 1.1\,Gyr. The reason for this age scaling is that a larger carbon
star contribution would mimic the spectrum of a younger cluster. The
arrows show, from top to bottom, an increasing fraction of carbon star
contribution to the final cluster light -- 40\,\%, 50\,\%, 60\,\% and
70\,\%, respectively originate from the C-type star. The same is valid
for both the CO index used by \citet{maraston05} and the \dco\/ index
defined by \citet{esther08}. This simple experiment shows us that we
are able to reproduce the lower observed CO index values in the
integrated spectra of LMC globular clusters having SWB type V with an
increasing fraction of the carbon star contribution. The same kind of
tests, but performed with integrated colours, make the cluster redder,
as expected when having an increased fraction of carbon rich stars.

The models of \citet{maraston05} include a careful treatment of the
TP-AGB stellar phase, which is of great importance for stellar
populations with ages between 0.3 and 2\,Gyr, due to its very high
luminosity. The empirical photometric calibration of the models has
been done with the near-IR photometric data of LMC globular clusters
and AGB stars of \citet{persson83} and \citet{frogel90}.  Spectra of
carbon-rich stars are also included. They come from the database of
\citet{lancon02}, which contains averaged C-type star spectra. These
authors have obtained 21 spectra of carbon-rich luminous pulsating variable stars in the Milky
Way. However, due to the small temperature scales of the sample, they
do not consider as justified to have more than a few averaged
bins. The carbon stars have been grouped according to their
temperature, defined by their $(H-K)$ colour. The first three bins
contain averages of 6 C-type star spectra each. Bins 4 and 5 contain
the spectra of a single very red star, R\,Lep, near maximum and
minimum light. The temperature of the stars is decreasing with
increasing bin number.

\begin{figure}
\resizebox{\hsize}{!}{\includegraphics[angle=0]{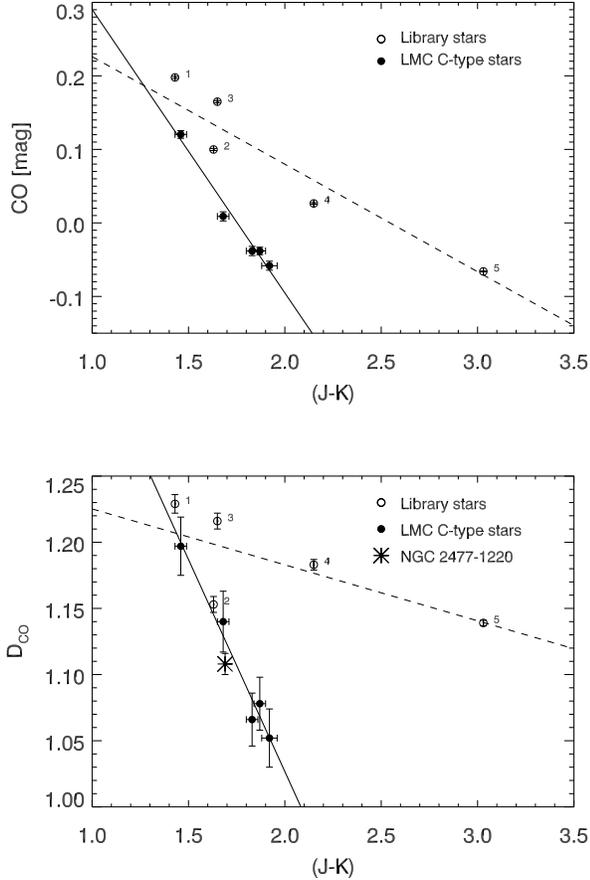}}
\caption{\label{fig:co_jk_cstars} $(J-K)$ vs. CO index in LMC and
  Milky Way carbon-rich stars. LMC stars are marked with filled
  circles. The linear least-squares fit to the data is shown with a
  solid line (Eq.~\ref{eq:fit_lmc}). The values for the Milky Way
  averaged star spectra from \citet{lancon02} are shown with open
  circles. The number next to each data point corresponds to the bin
  number. Their linear least-squares fit is shown with a dashed line
  (Eq.~\ref{eq:fit_mw}).}
\end{figure}

If carbon rich stars in the Galaxy and the LMC have different CO
absorption strengths, then this would have a profound impact on the
model predictions. Differences in the CO index have been observed by
e.g.  \citet{frogel90} using narrow-band filters, where they compare
the CO value of LMC clusters C-type stars and their counterparts in
the Milky Way. The LMC stars have systematically weaker CO indices at
a given colour. In Fig.~\ref{fig:co_jk_cstars} we show the CO index
values as a function of $(J-K)$ colour, measured in C-type stars
belonging to our sample (filled symbols), compared to the CO index
that we measured on the averaged spectra of \citet{lancon02}. The
least-squares linear fit to the LMC carbon rich stars is shown as a
solid line:

\begin{equation}
\label{eq:fit_lmc}
CO = 0.676\pm0.026 - 0.386\pm0.015\,(J-K)
\end{equation}
\begin{equation}
D_{CO} = 1.668\pm0.103 - 0.321\pm0.058\,(J-K)
\end{equation}

The Milky Way averaged spectra are marked with open symbols, with a
bin number assigned to each. The dashed line is showing the linear
least-squares fit to the averaged spectra:

\begin{equation}
CO = 0.373\pm0.002 - 0.146\pm0.001\,(J-K)
\end{equation}
\begin{equation}
\label{eq:fit_mw}
D_{CO} = 1.268\pm0.007 - 0.042\pm0.003\,(J-K)
\end{equation}

From this figure we see that the trends for C-type stars in the Milky
Way and the LMC generally disagree. Moreover, stars with $(J-K)>1.6$
have increasingly different CO index values, which explains why our
LMC GCs sample fits the model predictions for integrated colours, but
not for the CO index.

We made another test of including carbon star light to the spectrum of
NGC\,2173, but this time using the spectrum of bin 3 of
\citet{lancon02}. The result is shown in
Fig.~\ref{fig:co_models_comp} with a red arrow. For this case
specifically the ratio of carbon star to the original spectrum was
6:4. Different ratios, as in the previous test, gave similar
values. We see that the resultant CO index value increases by adding
Milky Way carbon stars and follows the trends predicted by the
models, while the inclusion of LMC cluster carbon stars leads to a
decrease of the index. We find this as an attractive explanation for
the discrepancy between the models and the data.
%
%

At this point the question arises why C-type stars in the LMC and the
Galaxy have different CO line strengths at a given near-IR colour? We
have to treat this issue with care, since not all of the carbon stars
in the \citet{lancon02} library disagree with the relation for the LMC
C-stars. The averaged spectrum from bin 2 agrees with the LMC
relation. In the bottom panel of Fig.~\ref{fig:co_jk_cstars} we
indicated with an asterisk another C-type star from the Milky Way,
located in the Galactic open cluster NGC\,2477 ([Fe/H]=\,--\,0.02,
$(J-K)=1.69$ \citep{houdashelt92}. A $K$-band spectrum of this star
has been published in \citet{silva2008}, however the authors did not
discuss its properties. Its colour and \dco\/ index value are
consistent with the relation for LMC C-type stars, given in
Equation~\ref{eq:fit_lmc}.

AGB stars in populations with lower metallicity are more likely to
become C-type. A star becomes C-type when the ratio of carbon to
oxygen atoms in its atmosphere becomes larger than one. Then all the
oxygen is locked up in CO molecules, and there is still some extra
carbon in the atmosphere. For the metal-poor stars there is less
oxygen in the atmosphere in the first place. Therefore these stars need
to dredge up less carbon in order to overcome, in relative amount, the
quantity of oxygen and so become C-type. The remaining carbon is used
to form other molecules, like C$_{2}$, CN and CH. In this way the CO
index as we measure it decreases. Note that the overall line strength
of the \co\/ may not decrease significantly. For example see
Fig.~\ref{fig:lmc_final_spec}, where the \co\/ features in NGC\,1806
and NGC\,2162 do not seem to be much weaker,  than in
NGC\,2173. But the CO indices, which we use to describe the behaviour
of this feature, take into account the continuum shape as well, which
in the case of carbon stars is severely affected by the presence of
features like C$_{2}$, CN or CH. In more metal rich stars the amount
of oxygen atoms is higher and all of the available carbon is used to
form CO molecules, which makes a stronger CO index in oxygen-rich stars. The different CO indices in Milky Way and LMC carbon-rich stars are due to a real change in the depth of the CO absorption features.

The observed difference between the \citet{lancon02} library carbon
stars in the Milky Way and the ones in our sample in the LMC has
important implications on the stellar populations models. The
photometric calibration of the intermediate age populations in the
models of \citet{maraston05} has been performed with globular clusters
in the LMC. However, for the spectral calibration, spectra of
carbon-rich stars in the Milky Way have been used, which leads to
inconsistent results. There might be a number of reasons for the
observed inconsistencies, like different metallicities of the carbon
stars or the phase of the pulsation in which they were observed. Our
results clearly show the necessity of a better understanding of the
properties of carbon rich stars and the need of larger empirical
stellar libraries to improve model predictions.

\section{Conclusions}
\label{sec:conclusions}

The goal of this project is to provide an empirical spectral library
in the near-IR for integrated stellar populations with ages $>$~1~Gyr,
which will be used to test the current and calibrate the future
stellar population models.  In this paper we have presented the first
results from a pilot study of the $K$-band spectroscopic properties of
a sample of six globular clusters in the LMC. To validate the
observational strategy, data reduction and analysis methods, we have
selected from the catalogue of \citet{bica99} three out of 38 GCs with
SWB type VII to represent the old ($>$10~Gyr) and metal poor
([Fe/H]$\sim$--1.4) population of the LMC, and three out of 71
clusters with SWB types V and VI to explore the properties of the
intermediate age (1\,--\,3~Gyr) and more metal rich
([Fe/H]$\sim$--0.4) component of the population. For each cluster our
integrated spectroscopy covers the central $24\arcsec\times24\arcsec$
and in most of the cases we have sampled about half the
light. However, in order to better sample bright AGB stars, which are
the most important contributors to the integrated cluster light in the
near-IR, we have observed up to 9 of the brightest stars outside the
central mosaics, but still within the tidal radii of the clusters,
that have near-IR colours and magnitudes consistent with bright red
giants in the observed clusters. We obtained integrated luminosity
weighted spectra for the six clusters, measured the line strengths of
\na\/, \ca\/ and \co\/ absorption features in the $K$-band and
compared the strength of \co\/ with the stellar population models of
\citet{maraston05}.

The observing strategy to cover at least the central half-light radius
with a number of SINFONI pointings showed to be an efficient way in
sampling the near-IR light of old ($>$\,10\,Gyr) clusters. For the
intermediate age and sparse clusters, which are dominated by just a
few very bright stars, observing a central mosaic plus a number of the
brightest stars in the vicinity of the cluster is a better choice for
optimal cluster light sampling. The availability of high spectral
resolution spectroscopy greatly helps the differentiation of the
cluster member stars from the LMC field population.

In intermediate age clusters the largest amount of light originates
from oxygen (M-type) and carbon-rich (C-type) AGB stars. Different
ratios of the contributions by these two types of stars can lead to
significant changes in the near-IR \co\/ line strength. According to
our observations, when the C-type stars contribution peaks (at
$\sim$1\,Gyr) the observed CO line strength is weak and then increases
fast to reach its maximum for clusters with age $\sim$2\,Gyr. It is
important to note that a weak line strength of \co\/ does not mean
that there is less CO in these clusters/stars. The indices, which are
used to describe the line strengths, take also into account the
continuum shape, which in the case of carbon-rich stars is severely
affected by typical for this type of stars absorption features and
thus the resultant index value is low.

The comparison of our data with the stellar population models of
\citet{maraston05}, in terms of CO line strength, shows a disagreement
for the youngest clusters in our sample. For clusters with age
$\sim$1\,Gyr the models predict maximal CO line strength, while we
observe the opposite: the CO strength is significantly weaker. At the same time, literature data of
the integrated colours of the clusters are consistent with these
models. We explain these discrepancies as due to the different origin of the C-type stars used
to calibrate the models and the ones in our data sample. The stars,
used for model calibration, are Milky Way carbon stars, while our
carbon stars are born in the LMC. We support this scenario with
Fig.~\ref{fig:co_jk_cstars}, where we show that carbon rich stars in
the Milky Way and LMC, which have similar $(J-K)$ colour, have very
different CO line strengths.  This deserves further investigation and hopefully the next generation of carbon star models \citep[e.g.][]{aringer09} will help us to elucidate whether a systematic effect of the metallicity on CO indices is expected, or the found  discrepancy is due to the small sample of individual observations of 
variable stars in existing libraries.
  
The near-IR \na\/ index shows a dependance on the age of the clusters
-- it is decreasing with an increasing age. The combination between
optical and near-IR spectral indices seems to offer possibilities to
break the age-metallicity degeneracy, but more accurate and detailed
stellar population models are necessary in the near-IR wavelength
range.

These models are of paramount importance, when studying the spatially
resolved stellar populations of nearby galaxies. Adaptive optics
assisted observations allow for the best correction of the Earth's
atmosphere perturbing effects when observing in the near-IR. First
attempts to explore galaxy evolution via spatially resolved near-IR
spectroscopy \citep[e.g.][]{az08,davidge08,nowak08} have shown the
need for a better understanding of the properties of stellar
populations in this wavelength range. The availability of detailed and
reliable stellar population models in the near-IR will open up a new
window to the exploration of galaxy formation and evolution.

 The final integrated spectra of the six LMC globular clusters will be made available via the Strasbourg astronomical Data Center (CDS).

\acknowledgements{We are grateful to many ESO staff astronomers who took the data presented in this paper in service mode operations at La Silla Paranal Observatory. We would like to thank Maria-Rosa Cioni, Claudia Maraston, and Daniel Thomas for helpful discussions. Finally, we thank the anonymous referee for her/his helpful suggestions.}

\bibliographystyle{aa}
\bibliography{/Users/mlyubeno/sci/biblio/papers,/Users/mlyubeno/sci/biblio/9314}

\end{document}